\documentclass[acmsmall]{acmart}
\AtBeginDocument{%
  \providecommand\BibTeX{{%
    \normalfont B\kern-0.5em{\scshape i\kern-0.25em b}\kern-0.8em\TeX}}}

\usepackage{hyperxmp}
\usepackage{dafny}
\usepackage{xspace}
\usepackage[inline,shortlabels]{enumitem}
\usepackage[T1]{fontenc}
\usepackage{booktabs}
\usepackage{algorithm}
\usepackage{algpseudocode}
\usepackage{xcolor}
\usepackage{fancyvrb}
\usepackage{subcaption}
\usepackage{graphicx}
\usepackage{amsmath}
\usepackage{tcolorbox}

\hypersetup{colorlinks,
  linkcolor=blue,
  citecolor=purple,
  urlcolor=blue,
  filecolor=blue}

\makeatletter
\renewcommand\paragraph{\@startsection{paragraph}{4}{0pt}%
  {-.2\baselineskip \@plus -2\p@ \@minus -.2\p@}%
  {-3.5\p@}%
  {\bfseries}}
\makeatother

\newcommand{\fstar}{\ensuremath{F^{*}}\xspace}

\newcommand{\T}[1]{\inlinedafny{#1}}

\newcommand{\redt}[1]{\textcolor{red}{#1}}

\newcommand{\grayt}[1]{\textcolor{gray}{#1}}

\newcommand{\naive}{na\"{i}ve\xspace}

\newcommand{\ie}{\emph{i.e.}\xspace}

\newcommand{\eg}{\emph{e.g.}\xspace}

\newcommand{\tname}[1]{\textsc{#1}\xspace}
\newcommand{\tool}{\tname{Laurel}}
\newcommand{\dataset}{\tname{DafnyGym}}
\newcommand{\cedar}{\tname{Cedar}}
\newcommand{\vmc}{\tname{DafnyVMC}}
\newcommand{\libraries}{\tname{Dafny-Libraries}}

\newcommand{\baseline}{\emph{Baseline}\xspace}
\newcommand{\errmsg}{\emph{Error}\xspace}
\newcommand{\iterative}{\emph{Iterative}\xspace}
\newcommand{\placeholder}{\emph{Placeholder}\xspace}
\newcommand{\placeholderr}{\emph{Placeholder+Error}\xspace}
\newcommand{\random}{\emph{Random}\xspace}
\newcommand{\similarity}{\emph{Similarity}\xspace}
\newcommand{\similaritynoplaceholder}{\emph{Similarity-no-placeholder}\xspace}
\newcommand{\embedding}{\emph{Embedding}\xspace}
\newcommand{\tfidf}{\emph{TF-IDF}\xspace}

\newcommand{\seqsim}{\ensuremath{\mathcal{S}}\xspace}
\newcommand{\tokensim}{\ensuremath{\mathsf{TokenSim}}\xspace}
\newcommand{\linesim}{\ensuremath{\mathsf{LineSim}}\xspace}
\newcommand{\codesim}{\ensuremath{\mathsf{ProofSim}}\xspace}

\newcommand{\ntasks}{143\xspace}
\newcommand{\maxattempts}{10\xspace}

\setcopyright{cc}
\setcctype{by}
\acmDOI{10.1145/3720499}
\acmYear{2025}
\acmJournal{PACMPL}
\acmVolume{9}
\acmNumber{OOPSLA1}
\acmArticle{134}
\acmMonth{4}
\received{2024-10-16}
\received[accepted]{2025-02-18}

\begin{document}

\title{\tool: Unblocking Automated Verification with Large Language Models}

\author{Eric Mugnier}
\orcid{0009-0006-4967-6820}
\affiliation{%
  \institution{UC San Diego}
  \city{San Diego}
  \country{USA}
}
\email{emugnier@ucsd.edu}

\author{Emmanuel Anaya Gonzalez}
\orcid{0009-0002-9013-2228}
\affiliation{%
  \institution{UC San Diego}
  \city{San Diego}
  \country{USA}
}
\email{fanayagonzalez@ucsd.edu}

\author{Nadia Polikarpova}
\orcid{0000-0001-5571-173X}
\affiliation{%
  \institution{UC San Diego}
  \city{San Diego}
  \country{USA}
}
\email{npolikarpova@ucsd.edu}

\author{Ranjit Jhala}
\orcid{0000-0002-1802-9421}
\affiliation{%
  \institution{UC San Diego}
  \city{San Diego}
  \country{USA}
}
\email{rjhala@ucsd.edu}

\author{Zhou Yuanyuan}
\orcid{0009-0006-0980-2538}
\affiliation{%
  \institution{UC San Diego}
  \city{San Diego}
  \country{USA}
}
\email{yyzhou@ucsd.edu}

\begin{abstract}

Program verifiers such as Dafny automate proofs by outsourcing them to an SMT solver.
This automation is not perfect, however,
and the solver often requires hints in the form of \emph{assertions},
creating a burden for the proof engineer.
In this paper, we propose \tool, a tool that
alleviates this burden by automatically generating assertions
using large language models (LLMs).

To improve the success rate of LLMs in this task,
we design two domain-specific prompting techniques.
First, we help the LLM determine the location of the missing assertion
by analyzing the verifier's error message and inserting an \emph{assertion placeholder} at that location.
Second, we provide the LLM with example assertions from the same codebase,
which we select based on a new \emph{proof similarity} metric.
We evaluate our techniques on our new benchmark \dataset,
a dataset of complex lemmas we extracted from three real-world Dafny codebases.
Our evaluation shows that \tool is able to generate over 56.6\% of the required assertions
given only a few attempts,
making LLMs an affordable tool for unblocking program verifiers without human intervention.
\end{abstract}


\maketitle

\section{Introduction}\label{sec:intro}

Auto-active verifiers like Dafny \cite{Leino2010Dafny},
$F^*$~\cite{fstar} and Verus~\cite{verus} significantly
automate the verification process, outsourcing low-level
details of proofs to an SMT solver~\cite{z3}.
Indeed, after decades of research, these verifiers
are getting adopted in industry, where they have been
used to build verified cryptographic
libraries~\cite{FHE2023Yang,AWSEncryptionSDK,evercrypt},
authorization protocols~\cite{Cedar},
a random number generator~\cite{dafnyVMC},
the Ethereum virtual machine~\cite{Ethereum2023Cassez},
packet parsers~\cite{everparse-fstar} and other critical
infrastructure.

\paragraph{The Problem: Assertion Hints}
Unfortunately, the reliance on SMT solvers is a double-edged sword,
as the automation provided by the solver is limited.
In practice, proof engineers often need to add hints
in the form of inline \emph{assertions} to their code
which introduce intermediate facts that break down the
proof obligation so that it can be readily digested
by the SMT solver.
For example, consider the following lemma
from the Dafny standard library~\cite{LibrariesDafny}:
\begin{dafny}
lemma LemmaCardinalityOfSet<T>(xs: seq<T>)
  ensures |ToSet(xs)| <= |xs|
{
  if |xs| == 0 {
  } else {
    assert ToSet(xs) == ToSet(DropLast(xs)) + {Last(xs)};
    LemmaCardinalityOfSet(DropLast(xs));
  }
}
\end{dafny}
The \emph{post-condition} of the lemma (\T{ensures})
states that the cardinality of a set generated from
a sequence \T{xs} is at most the length of \T{xs};
the body of the lemma constitutes its \emph{proof}.
The base case of the proof is discharged automatically,
while the case of a non-empty sequence makes
a recursive call in order to invoke the
induction hypothesis.
However, even with the induction hypothesis in scope
---- obtained by the post-condition or \emph{ensures}
clause of the recursive call --- the reasoning required
to prove the recursive case is too complex for the solver
to handle automatically.
Instead, the proof engineer must explicitly provide
a hint via the \emph{assertion} on Line 5 that explains
how to decompose the elements of a sequence if we split
off its last element, after which the solver can check
the proof.

\paragraph{Assertions are Difficult and Ubiquitous}
A significant amount of time in proof engineering is spent
on the tedious and time-consuming task of debugging failing
proofs --- including both where the verifier rejects the proof,
or where the verifier times out --- to divine the right assertion
hints that will guide the solver to the correct proof.
The SMT solver is a black box, making it hard to determine what
exact information is missing and needed to complete the proof,
or what extraneous information is making the solver timeout.
Indeed, a recent report on industrial verification for
cryptography~\cite{FormallyVerifyingIndustryCryptographyDodds}
reports that ``most proof engineering time is spent working with
a failing proof'' as opposed to, say, writing top-level function
specifications.
While other kinds of hints can also be used to debug
a failing proof~\cite{IroncladChris,VerificationOptimization},
we find that inline assertions are by far the most common,
and are ubiquitous in industrial proof efforts.
For example, assertions are the very first tactic listed
in the official Dafny proof optimization guide~\cite{VerificationOptimization}.
Moreover, we show~\autoref{sec:benchmark},
that in large industrial Dafny codebases
assertions comprise 4-11\% of the lines
of code, greatly exceeding other types
of hints from the guide
(see \autoref{sec:benchmark} for details).


In this paper, we present \tool: a tool that
helps proof engineers with the tedious task
debugging failing proofs by using Large Language
Models (LLMs) to automatically generate the assertions
needed to verify industrial Dafny code.
We develop \tool via the following concrete contributions.

\paragraph{1. Dataset Curation}
Our first contribution is \dataset, an open-source
dataset of Dafny lemmas extracted from three real-world
codebases: the Dafny standard library~\cite{LibrariesDafny},
an authorization library \cedar~\cite{Cedar}, and a library
of Monte-Carlo algorithms \vmc~\cite{dafnyVMC}.
Our analysis of \dataset shows that the code
in industrial proof developments are substantially larger
and more complex than the individual ``textbook'' procedures
considered by prior work applying LLMs to generate entire
proofs (\ie function bodies) from specifications (\ie function contracts)
~\cite{DafnyFSE24,sun2024clover,brandfonbrener2024VerMCTS,chakraborty2024neuralsynthesissmtassistedprooforiented,yang2024autoverus}.
In contrast, the proofs in our dataset make heavy use
of user-defined types, functions, and auxiliary lemmas
scattered throughout the codebase; all of which are
features that are out of scope for current whole-proof
generation approaches.
Thus, while the above work might seem to subsume the problem
of assertion synthesis, in fact as they were designed for and
evaluated on short, standalone programs, which are not
representative of industrial verification efforts, they
do not apply to the industrial code bases that we focus on.
As a concrete example, while much of the previous
work has targeted the automatic generation of loop
invariants (including using LLMs~\cite{kamath2023Invariant,loughridge2024dafnybench}),
our analysis reveals that there are in fact
\emph{vanishingly few} loop invariants in
the industrial codebases we analyzed---under 0.5\% of LOC ---
as large verification efforts mostly
contain lemmas and not executable code,
and lemmas rarely involve loops.
Consequently, \dataset shows that for large-scale verification
efforts, it is important to focus on the more specialized
problem of debugging an existing proof by generating proof hints.
Sadly, we found that with a na\"ive prompting strategy,
LLMs struggle to generate correct assertions for real-world
Dafny code.

\paragraph{2. Prompt Specialization}
Our second contribution is to improve the success rate of LLMs via two
key observations that we use to specialize prompts with
the necessary context.
First, we observe that LLMs often fail
to determine even the correct \emph{location} where
the assertion should be inserted.
Thus, our first prompt specialization technique
is a static analysis that uses the Dafny error
message to systematically infer the assertion
location, after which we can prompt the
LLM with an assertion placeholder for
that specific location.
Second, we observe that in most practical code evolution
scenarios, \eg when repairing proofs that have been ``broken''
by changes to the solver, or extending a mature codebase,
the codebase \emph{already contains} other assertions
similar to the one we are trying to generate.
Hence, we can use these existing assertions
as \emph{in-context examples}~\cite{fewshot} when
prompting the LLM.
The challenge, however, is that a codebase
may contain hundreds of assertions,
and giving them all to the LLM would
be detrimental to its performance.
We solve this problem with a novel, lightweight
language-agnostic \emph{hierarchical edit distance} metric
that captures the structural similarity of proofs.
We use this metric to design a heuristic that selects a small
set of assertions from lemmas that are structurally
similar to and hence, relevant to the target lemma, and
which can then be used to further specialize the LLM prompt
to more accurately generate the missing assertion.

\paragraph{3. Evaluation}
Our final contribution is an implementation of our approach
in \tool, a system that generates assertions for Dafny code
bases, and a comprehensive evaluation of \tool on the \dataset.
(Although \tool focuses on Dafny, the underlying
methods should readily transfer to other SMT-based
verifiers~\cite{key-book,whyml,fstar,LH,verus,creusot,prusti,flux}
that also rely on assertions to guide the solver.)
Our evaluation shows that, by specializing
the prompt locations, assertion placeholders
improve the success rate of LLMs by at least
a factor of 6 given ten attempts, compared
to a baseline prompting technique.
Further, the addition of in-context examples
selected via our proof similarity metric
improves the success rate
by up to 30\%.
Overall, \tool correctly generates 82/145 (56.6\%)
of the assertions in our dataset, showing that our
approach is viable for unblocking automated verification.

\section{Overview}\label{sec:overview}

In this section we illustrate the process of generating assertions in \tool,
using a simplified lemma from the \cedar codebase~\cite{Cedar} as a motivating example.
\begin{figure}[ht]
\begin{dafny}
lemma ParseDigitsAndDot(s1: string, s2: string)
  requires forall i | 0 <= i < |s1| :: '0' <= s1[] <= '9'
  ensures ParseDecStr(s1+"."+s2).value.0 == s1 && ParseDecStr(s1+"."+s2).value.1 == "."+s2
  {
    if |s1| == 0 {
    } else {
      ParseDigitsAndDot(s1[1..],s2);
    }
  }
\end{dafny}
\begin{dafnyNoLines}[basicstyle=\footnotesize\ttfamily,xleftmargin=0pt,xrightmargin=0pt]
(*@Error: a postcondition could not be proved on this return path@*)
      } else {

(*@Related location: this is the postcondition that could not be proved@*)
ensures ParseDecStr(s1+"."+s2).value.0 == s1 &&
        ^^^^^^^^^^^^^^^^^^^^^^^^^^^^^^^^^^^^
        ParseDecStr(s1+"."+s2).value.1 == "."+s2
\end{dafnyNoLines}
\caption{Example of a Dafny lemma that fails to verify without an assertion and the corresponding error message.}\label{fig:error}
\end{figure}

Consider the lemma \T{ParseDigitsAndDot} in \autoref{fig:error},
which specifies a correctness property of the function \T{ParseDecStr}:
namely, given an input string of the form \T{s1+"."+s2},
where \T{s1} contains only digits,
the function should return a pair \T{(s1, "."+s2)}
(\eg \T{ParseDecStr("123.456")} should return \T{("123", ".456")}).%
\footnote{The actual \cedar lemma requires that \finlinedafny{s1} be non-empty,
but we elide this requirement for exposition purposes.}

The programmer decides to prove this lemma by induction over the length of \T{s1};
to this end, they introduce a conditional to separate the base case from the inductive case,
and in the latter case, make a recursive call on the tail of \T{s1} to invoke the induction hypothesis.
Unfortunately, this is not enough to convince the Dafny verifier,
which returns the error message shown at the bottom of the figure.
According to this error message,
the solver is unable to prove that (the first conjunct of) the post-condition in Line~2
is established by the \T{else} branch (\ie the recursive case).
The reason for this error is not that the lemma is incorrect,
but rather that the chain of reasoning steps required to deduce the postcondition from the induction hypothesis is too long,
and the solver gets stuck along the way.
We can help the solver by providing an intermediate ``checkpoint'' that helps break down the proof;
more precisely, we need to insert the following \emph{assertion} at the end of the \T{else} branch (after Line~6):
\begin{equation}\label{eq:helper}
\mbox{\T{assert s1+"."+s2 == [s1[0]]+(s1[1..]+"."+s2)}}
\end{equation}
This assertion explains the relationship between the string passed to \T{ParseDecStr} in the current context
and its version passed to the recursive call.
It is challenging for the programmer to come up with this assertion
(unless they have extensive experience with similar proofs);
specifically, the error message provides very little insight into why the solver got stuck.

\paragraph{Can LLMs Generate Assertions?}

Encouraged by the success of LLMs in program repair~\cite{khatry2023words,ProgramRepairXia2023}
and proof synthesis~\cite{BaldurFSE23First,DafnyFSE24,kamath2023Invariant},
we conjecture that LLMs can be used to generate assertions automatically.
Unfortunately, na\"ively prompting an LLM with the code snippet and error message in \autoref{fig:error},
and asking it to fix this code by inserting an assertion does not work well.
Here are three sample assertions this \naive baseline yields in our experiments:
\begin{enumerate}[left=0pt]
  \item \T{assert forall i | 0 <= i < |s1| :: '0' <= s1[i] <= '9'} after Line~3;
  \item \T{assert '0' <= s1[0] <= '9'} after Line~4;
  \item \T{assert s2 != ""} after Line~5.
\end{enumerate}
Neither of the three assertions is helpful in proving the lemma
(and the last assertion also does not hold),
and moreover, they are not even inserted at the correct location!
We find this last result quite surprising,
since the placement of assertions---unlike their content---is usually easy to determine for human programmers
based on the error message.

\paragraph{The \tool Approach}

\begin{figure*}
  \centering
  \includegraphics[width=1\textwidth]{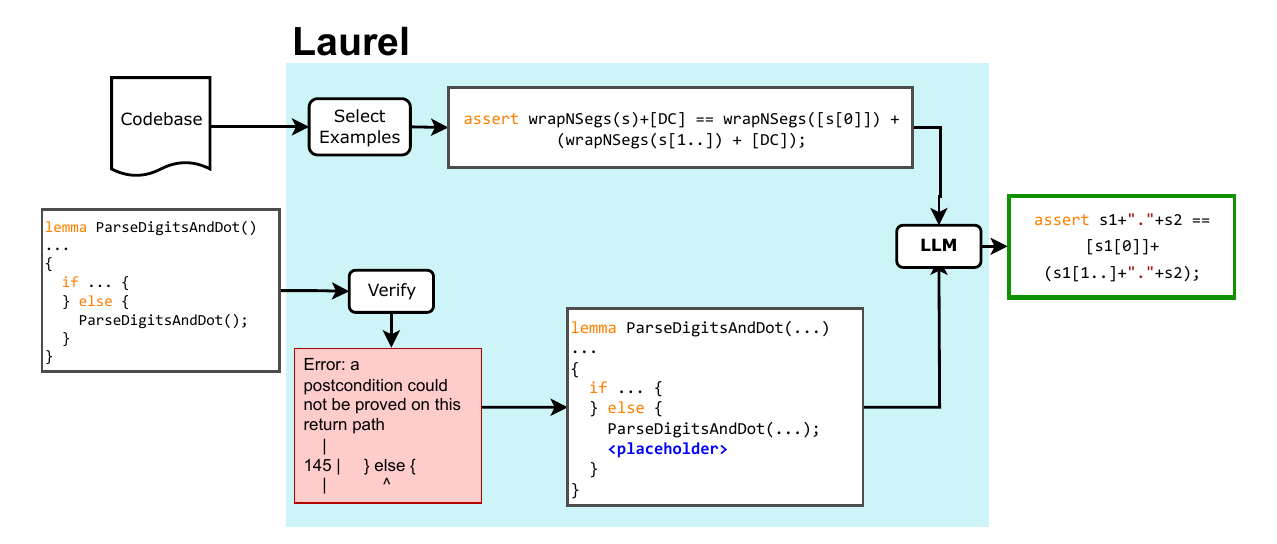}
  \caption{Overview of \tool: it takes two inputs,
  the lemma to fix, and the codebase used to find \emph{in-context examples}.
  It identifies the assertion location in the lemma’s \T{else} branch using
  the error message and selects relevant examples with a similar recursive structure.
  Finally, it prompts the LLM to generate an assertion.}
  \label{fig:overview}
\end{figure*}

To improve the success rate of LLMs in generating assertions,
we propose to augment the prompt with two crucial pieces of information:
\begin{enumerate*}
  \item the most likely \emph{location} where the assertion should be inserted;
  \item a set of relevant \emph{example assertions} from the same codebase.
\end{enumerate*}
We implement this approach in a tool called \tool,
whose high-level overview is shown in \autoref{fig:overview}.

\tool starts with two inputs:
\begin{enumerate*}[label=\alph*)]
  \item a Dafny codebase
  \item the name of the target lemma.%
  \footnote{A verification unit in Dafny is either a \emph{lemma} or a \emph{method}, depending on the presence of executable code;
  we refer to both as ``lemma'' for simplicity.}
\end{enumerate*}
The goal is to insert an assertion into the target lemma so that it verifies.
In this work, we assume that the lemma can be fixed by inserting a single assertion;
\autoref{sec:eval:threats} discusses how our approach might be extended to handle multiple assertions.

\tool constructs the LLM prompt in two key steps.
First, it runs the Dafny verifier on the target lemma to obtain the error message.
It then analyzes the error message to determine the most likely location for the assertion;
this location is used to insert an \emph{assertion placeholder} into the lemma,
as shown in \autoref{fig:overview}.
Second, \tool analyzes the codebase to identify a small set of $m$ relevant assertions
($m=6$ in our experiments)
to be used as \emph{in-context examples} (also called few-shot demonstrations~\cite{fewshot}).
\tool then constructs the prompt, including the lemma with the assertion placeholder and the example assertions,
and asking the LLM to generate an assertion to be inserted at the placeholder location.

Given an output from the LLM,
\tool replaces the placeholder with the generated assertion and re-runs the Dafny verifier;
if the target lemma still fails to verify,
\tool re-queries the LLM with the same prompt up to a fixed number $k$ of attempts
($k=10$ in our experiments).%
\footnote{In our experiments, to save cost, we sample $k$ outputs from the LLM at once, instead of querying the LLM $k$ times.}
The rest of this section gives a brief overview of our techniques for placeholder localization and example assertion selection;
\autoref{sec:placeholder} and \autoref{sec:lemma_similarity} provide more details.

\paragraph{Localizing the Assertion Placeholder}

Recall that with a \naive prompt,
the LLM struggles to determine the correct location for the missing assertion in our motivating example.
Our \textbf{first key insight} is that we can help the LLM localize the assertion by analyzing the error message.
Specifically, the error message in \autoref{fig:error} points to a \emph{violation} at the very end of the lemma
(which is where post-conditions are checked);
it also points to the \emph{faulty branch}---the \T{else} branch---which led to the violation.
Given this information, the optimal location for the assertion is at the end of the \T{else} branch:
placing it outside of the conditional would make assertion (\ref{eq:helper}) ill-formed
(since \T{s1} might be empty, so \T{s1[0]} might not exist);
on the other hand, placing it further away from the violation (\eg before the recursive call)
might not be as helpful, as at that point the solver is busy with something else (proving the preconditions of the recursive call).
\tool performs this analysis based on the control-flow graph of the lemma,
as we explain in \autoref{sec:placeholder}.

\paragraph{Selecting Example Assertions}

Once we know \emph{where} to insert the assertion,
the next challenge is to decide \emph{what} it should say.
Sometimes the missing assertion is syntactically similar to the failing specification
or to other assertions within the same lemma.
In these cases, the LLM can often generate the correct assertion
by taking hints from the error message or from the lemma itself.
This is not always the case, however,
and it is not the case in our motivating example.

Our \textbf{second key insight} is that the missing assertion is likely similar to other assertions
elsewhere in the same codebase,
which we can make available to the LLM as \emph{in-context examples}.
The challenge, however, is to select a small number of examples
that are \emph{relevant} to the current context
(that is, are likely to be syntactically similar to the missing assertion).
We further conjecture that relevant example assertions are likely to appear in proofs that are similar to our failing proof.
For instance, in our motivating example,
we might consider other lemmas that involve string manipulation
and have the structure of an inductive proof (with a base case and a recursive case).

To find such similar lemmas, \tool introduces a code similarity metric,
based on a hierarchical generalization of Levenshtein distance (\autoref{sec:lemma_similarity}),
which is effective at capturing structural similarity between code snippets.
\tool then selects $m$ assertions from the codebase whose context---%
\ie the surrounding lemma with the assertion removed---%
is most similar to our target lemma.

In our motivating example,
this algorithm recommends the following assertion from a lemma with a similar recursive structure:
\begin{equation*}
  \resizebox{\textwidth}{!}{\T{assert wrapNumSeg(s)+[DC] == wrapNumSegs([s[0]]) + (wrapNumSegs(s[1..]) + [DC])}}
 \end{equation*}
You can see that this assertion is syntactically similar to the missing assertion (\ref{eq:helper}).
Given this assertion as an example,
together with the assertion placeholder,
the LLM can now generate the correct assertion for the motivating example.
\section{Localizing Missing Assertions}\label{sec:placeholder}

To help the LLM localize the assertion,
\tool finds a potential location for the missing assertion in a failing proof.
As a running example throughout this section, we use the lemma in \autoref{fig:union-size},
where both post-conditions are violated,
unless an assertion is added inside the \T{if} branch in lines 7--10.

\begin{figure}
\begin{dafny}
lemma LemmaUnionSize<X>(xs: set<X>, ys: set<X>)
    ensures |xs + ys| >= |xs|
    ensures |xs + ys| >= |ys|
{
    if ys == {} {
    } else {
        var y :| y in ys;
        if y in xs {
            var xr := xs - {y};
            var yr := ys - {y};
            LemmaUnionSize(xr, yr);
        } else {...}
}}
\end{dafny}
\begin{dafnyNoLines}[basicstyle=\footnotesize\ttfamily,xleftmargin=0pt,xrightmargin=0pt]
(*@Error: a postcondition could not be proved on this return path@*)
    if y in xs {
(*@Related location: this is the postcondition that could not be proved@*)
    ensures |xs + ys| >= |xs|
(*@Error: a postcondition could not be proved on this return path@*)
    if y in xs {
(*@Related location: this is the postcondition that could not be proved@*)
    ensures |xs + ys| >= |ys|
\end{dafnyNoLines}
\caption{A failing Dafny lemma with complex control flow.}\label{fig:union-size}
\end{figure}

\begin{figure}
\includegraphics[width=.6\columnwidth]{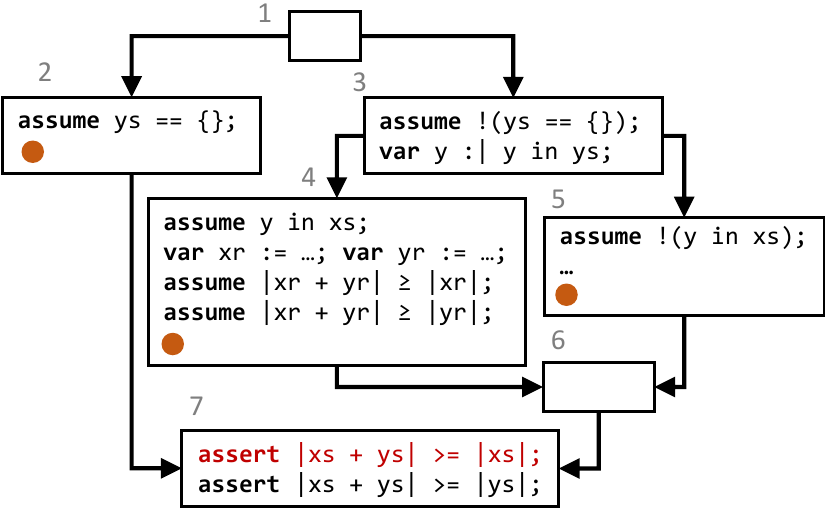}
\caption{Control flow graph (CFG) of the lemma in \autoref{fig:union-size}.
The highlighted \T{assert} corresponds to the first violation from the error message.
Orange circles denote candidate placeholder locations, \emph{were the faulty branch absent from the error message}.}\label{fig:cfg}
\end{figure}

At a high level, the localization proceeds in three steps:
\begin{enumerate*}[label=(\arabic*)]
\item \emph{desugaring} the Dafny program into an intermediate representation;
\item \emph{analyzing the error message} to identify the violations and the faulty branch; and
\item \emph{determining the placeholder locations} based on the control flow of the lemma and the information extracted from the error message.
\end{enumerate*}

\paragraph{Intermediate Representation}

To simplify the analysis, \tool operates at the level of intermediate representation
that Dafny uses to generate verification conditions~\cite{LEINO2005209,leino2008boogie}.
In this representation, the body of the lemma is represented as an acyclic \emph{control flow graph} (CFG),
and all specifications, calls, and loops are desugared into primitive \T{assume} and \T{assert} statements.
For example, the lemma in \autoref{fig:union-size} is desugared into the CFG shown in \autoref{fig:cfg};
here, the \T{assert} statements in node~7 of the CFG correspond to the post-conditions of the original lemma,
and instead of the call in line~10, node~4 simply assumes its post-conditions
(if the call had a pre-condition, it would first be asserted).

\paragraph{Error Message Analysis}

We formalize a Dafny error message as a sequence of \emph{violations},
where each violation is associated with an \T{assert} statement in the CFG that could not be proved.
In our example, the error message in \autoref{fig:union-size} contains two violations,
each associated with one \T{assert} statements in basic block~7 of the CFG.
\tool handles each violation separately, in the order they appear in the CFG;
for the rest of this section let us focus on the first violation.

For some violations,
the error message also identifies the \emph{faulty branch},
which corresponds to a node in the CFG on the path from the entry to the violation--node 4 in our example.

\paragraph{Placeholder Locations}

Given a violation, \tool determines the set of candidate locations for the assertion placeholder,
based on the following considerations:
\begin{enumerate}[left=0pt]
\item the placeholder should be \emph{as close as possible before} the violation in the CFG;
this allows the assertion to have the most influence on the solver when it tries to prove the violated assertion;
\item if the enclosing CFG node of the violation has \emph{ancestors that contain additional assumptions or local variables},
locations inside those ancestors should also be included,
because assertions placed inside those nodes have access to the additional assumptions and variables;
\item in the presence of the \emph{faulty branch} node,
only locations inside that node should be included.
\end{enumerate}

Returning to our example, if we assume that the error message does not identify the faulty branch,
candidate placeholder locations are denoted with orange circles in \autoref{fig:cfg}.
They are located in nodes~2, 4, and 5 because those are the ancestors of node~7 with additional assumptions and variables;
the candidate locations are at the end of their respective nodes,
since this makes them closer to the violation and also ensures that they can benefit from all the assumptions and variables in the node.
Once \tool takes into account the faulty branch information, however,
the only remaining candidate location is at the end of node~4.
Indeed, adding the assertion \T{assert xr + yr == xs + ys - {y};} at this location fixes the violation;
note that this assertion uses the local variables \T{xr}, \T{yr}, and \T{y} that are in scope at this location,
but not, \eg, inside node~7.

\section{Selecting Examples from Similar Proofs}\label{sec:lemma_similarity}

\tool selects relevant in-context examples to prompt the LLM.
The inputs to this task are
\begin{enumerate*}[label=(\arabic*)]
    \item a target \emph{proof context} $C^*$, where a proof context is a lemma with a single placeholder for the missing assertion, and
    \item a collection of \emph{demonstration examples} $\mathcal{E} = \{(C_i, a_i)\}_{i}$,
    where each $C_i$ is a proof context, and $a_i$ is its target assertion (\ie the assertion that repairs the proof).
\end{enumerate*}
Our goal is to select $m$ examples from $\mathcal{E}$ that are the most helpful to repair $C^*$.

Notably, the most useful demonstration examples are commonly those that contain syntactically similar assertions
to the target assertion. Crucially, it is often the case that \emph{similar assertions often occur within similar proof contexts}.
The approach then is to select those demonstration examples $(C_i, a_i)$ where $C_i$ is the most \emph{similar} to $C^*$,
with the intuition that the assertions $a_i$ will be also be similar to our target assertion $a^*$,
and therefore help the LLM generate $a^*$.
The problem then turns into defining an effective similarity notion between proof contexts.

Measuring source code similarity is a problem that arises in many areas of computer science research,
and as such, extensive literature exists on the topic \cite{zakerinasrabadi2023similarity}.
Among the most accessible and widely used techniques are TF-IDF and Neural Embedding vector search
\cite{karpukhin-etal-2020-dense,leandojo,neelakantan2022textcodeembeddingscontrastive,mikula2023magnushammer}, which rely on computing a vector representation of documents---using word frequency statistics and neural networks, respectively---
for both the query program and a bank of example programs. A relatedness measure, such as cosine similarity, is then used to find the most relevant element given the query.
However, these embedding-based techniques overlook important aspects of code similarity. Specifically,
they fail to consider the structure of source code and the ways programmers interact with it,
and are not robust to code-specific transformations, such as variable renaming.

Given these limitations, one can turn to tree-based similarity metrics,
which instead determine similarity by comparing the syntax trees of the programs.
Although these techniques address both limitations mentioned above, they introduce challenges of their own:
first, creating parse trees for a large codebase can be implementation-heavy and time-consuming.
Additionally, parsers for mainstream languages are often complex and difficult to adapt to specific use cases,
such as parsing and comparing incomplete programs.
Furthermore, matching subtrees is computationally expensive;
even the simplest tree edit distance algorithms have a time complexity that is worse than quadratic \cite{zhang_shasha_1989}.
Finally, most state-of-the-art tree-based similarity tools are only implemented
and ad-hoc optimized for a handful of popular programming languages, making them unsuitable for
many scenarios, including the one we currently face.

An effective similarity algorithm would ideally achieve the performance of tree-based techniques,
while retaining the flexibility of text-based comparison methods.
Here, we propose a similarity algorithm that takes into account the hierarchical structure of code without requiring parsing.
We leverage the organization of imperative code as a \emph{sequence of lines},
which can in turn be logically represented as a \emph{sequence of lexer tokens} of the language,
and ultimately as \emph{sequences of characters}.
Furthermore, we observe that the design of many programming languages and programming practices show a tendency toward \emph{aligning}
the different parts of the syntax tree with this natural hierarchy (see \autoref{fig:code_sim_target}, \autoref{fig:code_sim_example}).
A similarity algorithm that leverages this structure can closely approximate the tree edit distance
without the need for language-specific parsing, thereby remaining applicable to a wider range of scenarios.

\paragraph{Proof Similarity Metric}

\begin{algorithm}
    \caption{Sequence similarity}\label{alg:seq-sim}
    \begin{algorithmic}[1]
        \Function{SeqSim}{$s_1, s_2, \sigma$}
        \State $n \gets |s_1|$
        \State $m \gets |s_2|$
        \State $dp \gets \text{2DArray}(n+1, m+1, 0)$
        \For{$i \gets 1 \dots n$}
            \For{$j \gets 1 \dots m$}
                \State $dp[i][j] \gets \max\{$
                \State \qquad $dp[i-1][j],$
                \State \qquad $dp[i][j-1],$
                \State \qquad$dp[i-1][j-1] + \sigma(s_1[i], s_2[j])$
                \State $\}$
            \EndFor
        \EndFor
        \State \Return $\frac{dp[n][m]}{\max\{n, m\}}$
        \EndFunction
    \end{algorithmic}
\end{algorithm}

Our core algorithm computes the similarity score $\seqsim_\sigma(s_1, s_2) \in [0, 1]$
between two sequences $s_1, s_2 \in \Sigma^*$ over an arbitrary alphabet $\Sigma$. A
score of $1$ indicates identical sequences, while a score of $0$ indicates no commonality.
The algorithm identifies an optimal matching between the elements of $s_1$ and $s_2$
using a Warren-Fischer style~\cite{wagner_fischer-1974-edit} dynamic programming approach;
the full details are shown in Algorithm \ref{alg:seq-sim}.
Our approach differs from the standard string edit distance in two key aspects.
First, the algorithm takes an additional parameter $\sigma\colon \Sigma \to [0,1]$,
which is a similarity function on the elements of $s_1$ and $s_2$.
The standard string edit distance is a special case
where $\sigma(a, b) = 1$ if $a = b$ and $0$ otherwise.
Second, our algorithm converts the distance $d$ between the strings
into a similarity score by normalizing it and subtracting from $1$:
$$
\seqsim_\sigma(s_1, s_2) = 1 - \frac{d}{\max(|s_1|, |s_2|)}
$$

A key property of this algorithm is that it can be used \emph{hierarchically}:
we can measure the similarity over \emph{sequences of sequences},
by passing $\seqsim$ as the $\sigma$ parameter at the next level.
For instance, \tool uses this property to define the similarity between proof contexts as
$\codesim(C_1, C_2) = \seqsim_\linesim(C_1, C_2)$,
where $C_1, C_2$ are interpreted as sequences of lines,
and $\linesim$ is the similarity between lines,
calculated using our algorithm:
$\linesim(l_1, l_2) = \seqsim_{\tokensim}(l_1, l_2)$.
Finally, $\tokensim$ is the similarity function between tokens,
manually defined to treat all identifiers as equal
but to consider all other tokens as equal only to themselves.

\begin{figure*}[t]
    \centering
    \begin{subfigure}[b]{1\textwidth}
        \begin{Verbatim}[commandchars=\\\{\},fontsize=\footnotesize, frame=lines]
\grayt{// a1}
\grayt{\redt{assert s1 +} "." + \redt{s2 =}} \grayt{\redt{[s1[0]] +}} \grayt{\redt{(s1[1..] + }"." + \redt{s2)}}

\grayt{// a2}
\grayt{\redt{assert} wrapNumSegs(\redt{s}) \redt{+} [\redt{DC}] \redt{=}} \grayt{wrapNumSegs(\redt{[s[0]]}) \redt{+}} \grayt{\redt{(}wrapNumSegs (\redt{s[1..]})\redt{+}[\redt{DC}]\redt{)}}
        \end{Verbatim}
        \caption{Line similarity computation}
        \label{fig:line_sim}
    \end{subfigure}
    \newline
    \newline
    \begin{subfigure}[b]{0.50\textwidth}
        \begin{Verbatim}[xleftmargin=12pt, numbers=left, numbersep=6pt, commandchars=\~\^\&, codes={\catcode`$=3},fontsize=\footnotesize, frame=lines]
~violett^lemma ParseDigitsAndDot(s1:string, s2:string)&
~grayt^requires forall i | 0 <= i < |s1|&
~grayt^  :: '0' <= s1[i] <= '9'&
~bluet^ensures ParseDecStr(s1+"."+s2).v.1 == "."+s2&
~magt^{&
~purplet^  if |s1| == 1 {&
~grayt^    assert ParseDecStr("."+s2).None?;&
~cyant^  } else {&
~grayt^    ParseDigitsAndDot(s1[1..],s2);&
~redt^    <assertion_here>&
~purplet^  }&
~magt^}&
        \end{Verbatim}
        \caption{Target proof context $C^*$}
        \label{fig:code_sim_target}
    \end{subfigure}
    \hfill
    \begin{subfigure}[b]{0.45\textwidth}
        \begin{Verbatim}[xleftmargin=12pt, numbers=left, numbersep=6pt, commandchars=\~\^\&, codes={\catcode`$=3},fontsize=\footnotesize, frame=lines]
~violett^lemma CountDCLast(s:seq<numV6>)&


~bluet^ensures countDC(wrapNumSegs(s)+[DC]) == 1&
~magt^{&
~purplet^  if |s| == 0 {&

~cyant^  } else {&

~redt^    <assertion_here>&
~purplet^  }&
~magt^}&
        \end{Verbatim}
        \caption{Example proof context $C$}
        \label{fig:code_sim_example}
    \end{subfigure}
    \caption{Similarity metric examples}
    \label{fig:similarity_examples}
\end{figure*}

\paragraph{Examples}

Let us first illustrate the line similarity function, \linesim.
As the two lines, consider the two assertions $a_1, a_2$ from \autoref{sec:overview}, shown in \autoref{fig:line_sim}.
Tokens matched by \linesim are highlighted in red,
while unmatched tokens are in gray, indicating they need to be inserted or deleted.
Because 22 out of 35 of the tokens in the longer second line are matched,
\linesim will give a high similarity score of $22/35 \approx 0.63$.

Next, consider \codesim,
with two proof contexts $C^*, C$ in \autoref{fig:code_sim_target} and \autoref{fig:code_sim_example},
where $C^*$ is the target context from \autoref{sec:overview} and $C$ is an example context from the codebase.
Lines matched by \codesim share colors,
while unmatched lines are in gray.
The algorithm accurately identifies structurally similar lines
using \linesim as the element similarity inside \codesim.
Since most of the lines are being matched, the similarity score will be high,
appropriately leading to selecting $C$---along with its omitted assertion, $a_2$ from \autoref{fig:line_sim}---as an in-context example for $C^*$.

\section{The \dataset Dataset}\label{sec:benchmark}

Since existing Dafny benchmarks~\cite{sun2024clover,DafnyFSE24,loughridge2024dafnybench}
are limited to textbook-style problems and standalone functions,
we create a new benchmark dataset for assertion synthesis, which we dub \dataset.%
\footnote{Our dataset is available as part of the supplementary material,
and will be made public upon acceptance.}

\paragraph{Codebases}

\begin{figure}
  \centering
  \includegraphics[width=0.6\linewidth]{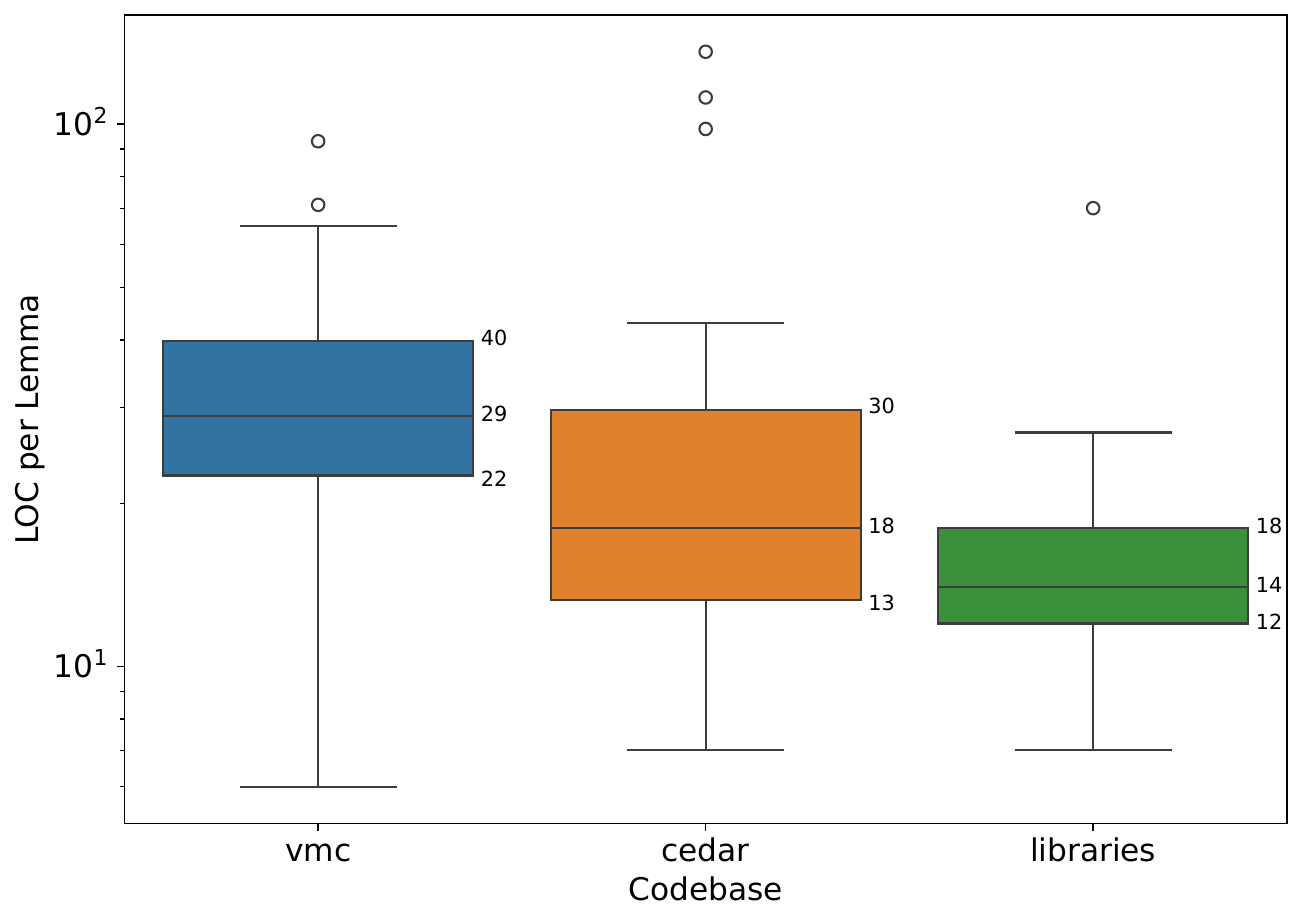}
  \caption{Number of lines per lemma for each codebase in \dataset.}
  \label{fig:codesize}
\end{figure}

We build the dataset from three Dafny codebases:
\begin{itemize}
  \item Dafny Libraries~\cite{LibrariesDafny} is
  a standard library of Dafny functions and lemmas
  useful for working with collections, non-linear arithmetic, file I/O, JSON, and more.
  \item \cedar Specification~\cite{Cedar} is a Dafny formalization of the Cedar authorization policy language used by AWS.%
  \footnote{A new version of Cedar has been released since the start of this project that uses Lean instead of Dafny.}
  \item \vmc~\cite{dafnyVMC} is a verified implementation of the Monte-Carlo algorithms and random number generators.
\end{itemize}
All three codebases
are maintained by professional software engineers at Amazon.
As shown in \autoref{tab:benchmark},
their size ranges from 4k to 18k lines of code,
containing between 219 and 2613 assertions which represent between
5.3\% and 10.8\% of the codebase.
Note that other hints such as loop invariants are barely used in these codebases
with only 108 loop invariants in total, representing 0.3\% of the lines of code.
The codebases are also diverse, targeting three different problem domains.

\begin{table}[]
  \caption{Codebases used to build \dataset}
\resizebox{\textwidth}{!}{%
  \begin{tabular}{@{}lrrr|r@{}}
  \toprule
                    & \multicolumn{1}{c}{Libraries} & \multicolumn{1}{c}{\cedar} & \multicolumn{1}{c}{\vmc} & \multicolumn{1}{c}{Total} \\
\midrule
  LOC               & 18,316    & 10,674 & 4,064 & 33,054 \\
  Assertions        & 1234  (6.7\% of LOC )     & 1160  (10.8\% of LOC)  & 219  (5.3\% of LOC)  & 2,613 (7.9\% of LOC) \\
  Loop invariants  & 99 (0.5\% of LOC)           & 0         & 9 (0.2\% of LOC)      & 108 (0.3\% of LOC)       \\
  \midrule
  Extracted assertions & 58        & 54    & 33    & 145    \\
  \bottomrule
  \end{tabular}%
}
  \label{tab:benchmark}
\end{table}

\paragraph{Extracting Benchmark Assertions}

To determine if an assertion is correctly generated by \tool,
we need an oracle to judge its correctness and relevance.
One option is to use the existing assertion as ground truth,
considering the generated assertion correct
if it is syntactically identical to the existing one.
However, this approach is limited, as
there are often multiple ways to express the same assertion,
and our main objective is to assist the verifier in proving the lemma,
not to replicate existing assertions.
Instead, we rely on the verifier itself as an oracle:
we consider an assertion correctly generated
if, and only if, the verifier fails to prove the lemma without it.
Not all assertions in the codebase are directly necessary for the verifier’s success.
To identify those that are essential,
we remove each assertion individually using a regular expression and re-verify the target lemma,
including the assertion into our dataset only if its removal causes the proof to fail.

Interestingly, we found a significant portion of assertions in these codebases were not actually necessary for proof completion
(though we caution that the table does not indicate all unnecessary assertions,
because if $n$ assertions from the same lemma can be removed one by one, it does not mean that all $n$ can be removed simultaneously).
While some of these unnecessary assertions prevent proof brittleness,
many seem to result from common development practices, where proof engineers try various assertions and may neglect to remove unneeded ones.
As these superfluous assertions can slow down verification,
 \tool offers a solution to automate assertion synthesis,
 ultimately producing cleaner, more efficient proofs.

\begin{figure}
\begin{dafny}[xrightmargin=0pt]
lemma SeqAddInequality<T>(s1: seq<T>, t1: T, s2: seq<T>, t2: T)
requires s1 != s2 || t1 != t2
ensures s1 + [t1] != s2 + [t2]
{
  if s1 == s2 {
    assert t1 != t2;
    var len := |s1|;
    assert (s1 + [t1])[len] != (s2 + [t2])[len];
  } else if |s1| == |s2| {
    var i :| 0 <= i < |s1| && s1[i] != s2[i];
    assert (s1 + [t1])[i] != (s2 + [t2])[i];
  }
}
\end{dafny}
\caption{A Dafny lemma with three assertions, out of which only two are \emph{assertions}, \ie necessary for the proof.}\label{fig:dataset-creation}
\end{figure}

Consider the Dafny lemma in \autoref{fig:dataset-creation},
which includes three assertions in Lines 4, 6, and 9.
We first remove the assertion in Line 4 and re-verify the lemma,
which leads to a verification failure;
therefore, this assertion/lemma pair is included in the dataset.
Next, we remove the assertion in Line 6,
but the proof succeeds with the remaining two assertions, so this assertion is excluded from the dataset.
Finally, we repeat the process for the assertion in Line 9,
which again leads to a verification failure, so it is included.

For each assertion,
we also check for exact duplicates within the same lemma.
If duplicates exist, we exclude the assertion from the dataset
to avoid overly simple synthesis tasks, where an identical assertion appears as an in-context example.

\paragraph{Assertions and Lemmas}
\autoref{tab:benchmark} shows the number of assertions extracted from each codebase,
yielding a total of \ntasks assertion synthesis tasks in \dataset.
This count represents only a fraction of the total assertions across the codebases,
as our regular expressions do not capture all possible assertion syntactic forms,
and duplicates are removed.
However, these assertions are integral to complex lemmas
that cannot be proven without them.

To assess the complexity of these assertions, we measured
the number of lines in the lemmas containing them as shown in \autoref{fig:codesize}.
Most lemmas exceed 10 lines of code,
with half spanning more than 18 lines of code,
indicating that they are non-trivial
and require significant effort to prove.

\section{Evaluation}\label{sec:eval}

We design our evaluation to answer three main research questions:
\begin{enumerate}[\bfseries RQ1)]
  \item How effective are LLMs with a baseline prompt at generating assertions?
  \item Do \emph{assertion placeholders} help LLMs generate Dafny assertions?
  \item Does selecting in-context examples via \emph{proof similarity} help LLMs generate Dafny assertions?
We also compare our \emph{proof similarity} metric with two state-of-the-art similarity metrics: \emph{neural embedding} and \emph{TF-IDF}.
\end{enumerate}

\subsection{Experimental setup}\label{sec:eval:setup}

To answer these research questions,
we run different variants of \tool, corresponding to different prompting techniques,
on the \ntasks assertion synthesis tasks from \dataset.
For each task, we give the LLM \maxattempts tries to generate an assertion,
and we consider the task solved in $k \leq \maxattempts$ attempts
if any of the first $k$ generated assertions causes the lemma to verify.
We then compare the variants based on the number of tasks solved for each $k$.
We do not compare \tool with other tools,
since to the best of our knowledge, there are no existing tools that target assertion generation.

We run all of our experiments on a server with a 2.83GHz Quad-Core Intel(R) Xeon(R) CPU X3363,
16 Gib of RAM and running Ubuntu 20.04.
We use Dafny 4.3.0 as the verifier and GPT-4o~\cite{GPT4o} as the LLM.
When prompting the LLM, we use standard best practices,
such as providing a system prompt,
as shown in \autoref{fig:prompt} (see appendix \autoref{appendix:prompt} for a full prompt example).
We set the temperature to $1.0$ to increase the variety of generations.

\begin{figure}
\begin{tcolorbox}[title={Prompt}]
\textbf{System instructions:}\\
You are a Dafny formal method expert.
You will be provided with a Dafny lemma that does not verify.
Your task is to insert an assertion in the <assertion> placeholder to make it verify.

\begin{tcolorbox}[title={In-context example extracted using \emph{proof similarity}}]
\medskip
\textbf{User: Example Task}\\
Can you fix this proof by inserting one assertion in the <assertion> placeholder?
\begin{verbatim}
lemma LemmaCardinalityOfSet<T>(xs: seq<T>) {
  ...
}
\end{verbatim}
\medskip
\textbf{AI Assistant: Example Solution}
\begin{verbatim}
  assert wrapNSegs(s)+[DC] == wrapNSegs([s[0]]) +
(wrapNSegs(s[1..]) + [DC]);
\end{verbatim}
\end{tcolorbox}

\begin{tcolorbox}[title={Lemma to fix with \emph{Placeholder}}]
\medskip
\textbf{User:}\\
Can you fix this proof by inserting one assertion in the <assertion> placeholder?
\begin{verbatim}

lemma ParseDigitsAndDot()
  ...
{
  if ... {
  } else {
    <assertion> Insert assertion here </assertion>
    ParseDigitsAndDot();
  }
}
\end{verbatim}
\end{tcolorbox}

\end{tcolorbox}
\caption{Example of prompt after integrating the results from \placeholder and \emph{Proof similarity}.}
\label{fig:prompt}
\end{figure}
\subsection{RQ1: Effectiveness of Baseline LLM Prompts}\label{sec:eval:rq0}

\begin{figure}
\centering
\includegraphics[width=1\linewidth]{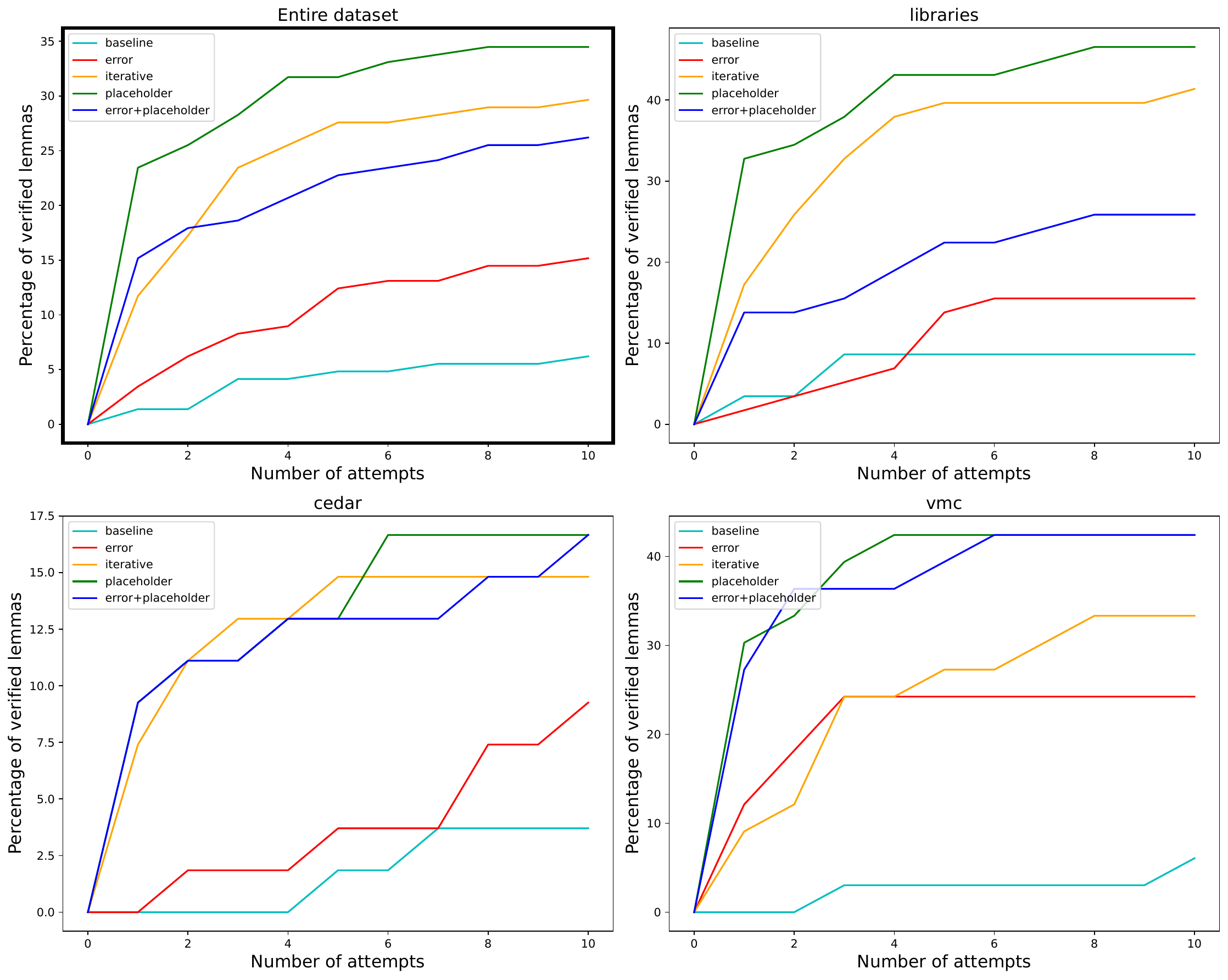}
\caption{Percentage of verified lemmas vs inference cost for the entire \dataset dataset, and by codebase, using placeholder.}
\label{fig:rq1_total}
\end{figure}

In this experiment, we assess the baseline effectiveness of the LLM in generating assertions;
namely, we evaluate the following three prompting strategies:
\begin{itemize}
  \item \baseline: the LLM is given only the code of the target lemma.
  \item \errmsg: the LLM is given the code and the initial error message produced by Dafny.
  \item \iterative: same as baseline \baseline,
  but if the lemma still does not verify with the generated assertion,
  we prompt the LLM again with the resulting error message.
\end{itemize}

\autoref{fig:rq1_total} presents the success rate of these three variants
(along with other variants, discussed in \autoref{sec:eval:rq1})
on the entire \dataset dataset
and also with the results divided by codebase.
Specifically, the x-axis shows the number of attempts $k = [1..10]$
and the y-axis shows the percentage of lemmas verified in $\leq k$ attempts.

As you can see in \autoref{fig:rq1_total},
\baseline performs quite poorly across the board,
achieving a success rate of 6.2\% even at $k=10$.
The LLM performs best on Libraries with 8.6\%,
and worst on \cedar with only 3.7\% of tasks solved.
The difference is consistent with the complexity of the codebases:
the lemmas in Libraries mainly deal with standard data structures and mathematical concepts,
while those in \cedar are domain-specific
and contain user-defined types and functions.

Manual analysis shows that the poor performance of \baseline
is due to the inability of the LLM to identify and localize the issue in the lemma.
As we illustrated in \autoref{sec:overview},
the LLM may attempt to add an assertion at an unrelated location,
or even try to modify a pre- or a post-condition.
Note that the LLM's confusion about the location of the assertion has a complex effect:
even a correct assertion inserted at a wrong location can lead to a verification failure,
but also the assertion itself is more likely to be incorrect,
because the surrounding context is different.

Including the raw Dafny error message in the prompt more than doubles the LLM's overall success rate,
but in absolute terms it remains low, at around 15\%:
the LLM seems to have difficulties linking the error message with the correct line in the lemma.

\begin{figure}
  \centering
  \includegraphics[width=.6\linewidth]{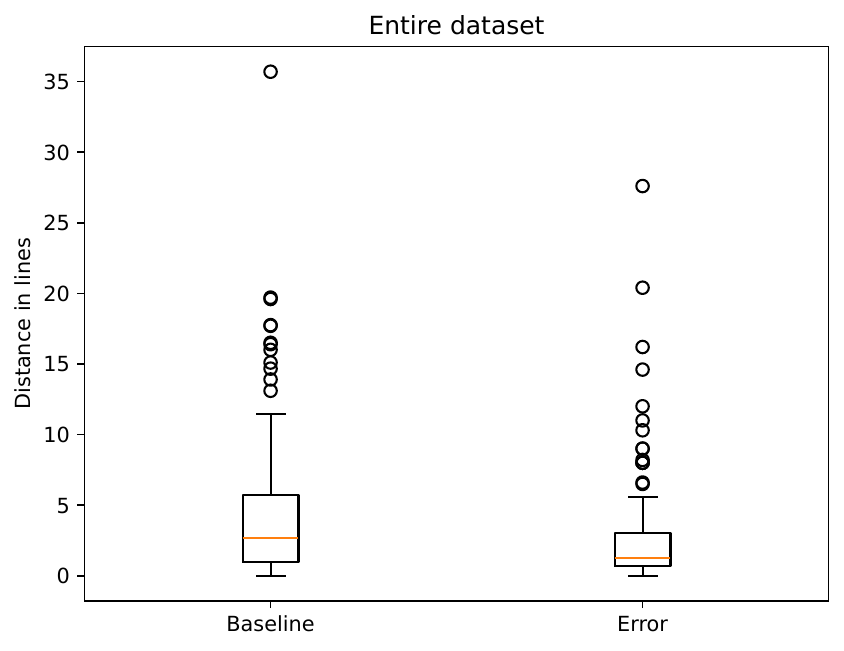}
  \caption{Distance to the LLM's chosen assertion location from the ground truth assertion on the entire \dataset dataset.}
  \label{fig:lines_distance}
\end{figure}

To confirm our hypothesis that the LLM struggles with assertion placement,
we measured the distance between the LLM's chosen assertion location and the ground truth location in the original codebase;%
\footnote{Although the ground truth location is not the only correct location for an assertion,
larger distances are indicative of the LLM's inability to localize the issue.}
the results are depicted in \autoref{fig:lines_distance}.
As suspected, \baseline struggles to find the correct location,
with a median distance of 2.8 lines, a third quartile of 5.85 lines, and many outliers over 13 lines away.
Adding the error message does improve the placement, but many assertions are still placed far from the ground truth.

Iterative prompting increases the success rate substantially to 29.6\% on the entire dataset.
As indicated by previous work~\cite{AutomatedProgramRepairXia2024Issta},
this suggests that the LLM has the ability to learn from its mistakes
when provided with feedback on incorrectly generated assertions.
Note, however, that iterative prompting effectively doubles the number of attempts the LLM gets to generate an assertion,
so the comparison is not entirely fair.

\subsection{RQ2: The Effect of Assertion Placeholders}\label{sec:eval:rq1}

For this experiment, we compare the baseline prompting strategies
from the previous section with the following two variants of \tool:
\begin{itemize}
  \item \placeholder: the LLM is given the code with an assertion placeholder.
  \item \placeholderr: the LLM is given the code with an assertion placeholder and the error message.
\end{itemize}
\autoref{fig:rq1_total} shows the success rate of these variants on the \dataset dataset,
and the results by codebase.

As you can see, \placeholder significantly improves the success rate over the baselines,
up to 34.4\% on the entire dataset for $k=10$.
Additionally, more than half of these assertions are generated on the first try.
We attribute this improvement to the fact that LLMs
are better at code in-filling than at error localization,
as reported by previous work on program repair~\cite{ProgramRepairXia2023}.

As for individual codebases,
once again \cedar has the lowest success rate with 16.6\%,
while Libraries and \vmc rise above 40\%.
This is consistent with our earlier observation that \cedar assertions are complex,
and are harder for the LLM to generate, even when the location is known.

Adding the error message to the placeholder does not enhance the overall success rate, which drops to 26.2\%.
Comparing across codebases, however, we can see that adding the error message is detrimental for Libraries,
while it does not impact the overall success rate for \cedar and \vmc.
Our intuition is that the error message can sometimes guide the LLM in identifying the problem to be addressed,
but it can also mislead it or inhibit the generation of diverse outputs.
This effect is visible even in \vmc and \cedar, where, although the error messages are helpful overall,
\placeholderr takes more tries to reach the maximum success rate,
because error messages tend to get the LLM stuck on reproducing the failing specification in the assertion.

\subsection{RQ3: The Effect of Proof Similarity}\label{sec:eval:rq2}

\begin{figure}
  \centering
  \includegraphics[width=1\linewidth]{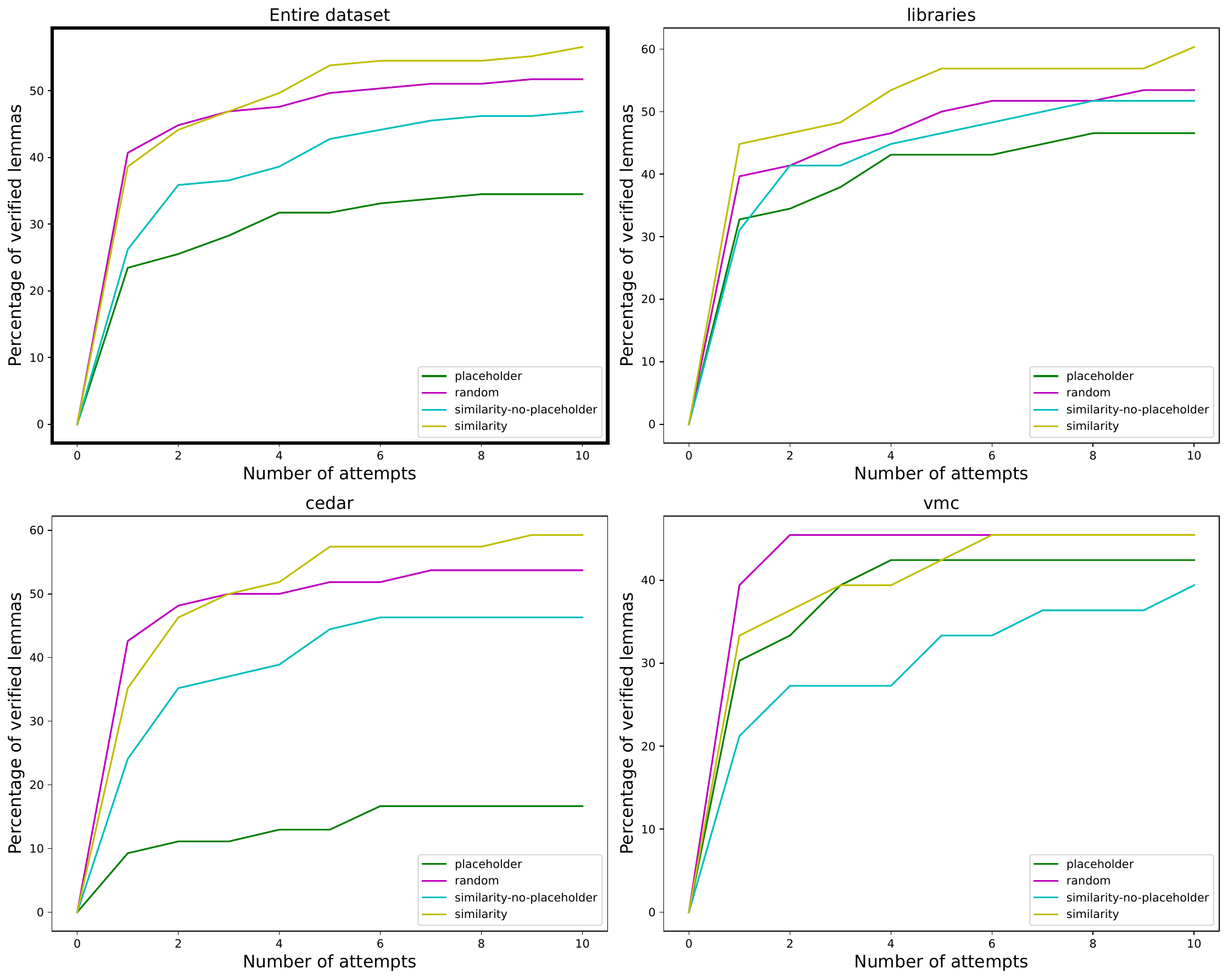}
  \caption{Percentage of verified lemmas vs inference cost for the entire \dataset dataset, and by codebase, using similarity.}
  \label{fig:rq2_total}
\end{figure}

In this experiment, we evaluate the effectiveness of selecting in-context examples
based on proof similarity,
by comparing the following variants of \tool:
\begin{itemize}
  \item \placeholder: the top-performing variant from RQ2.
  \item \random: the LLM is given the code with a placeholder and $m$ examples from the same codebase, selected at random.
  \item \similarity: the LLM is given the code with a placeholder and $m$ examples from the same codebase, selected based on proof similarity.
  \item \similaritynoplaceholder: the LLM is given the code without placeholder and $m$ examples from the same codebase, selected based on proof similarity.
\end{itemize}
We use $m=6$ in all experiments.
For both \random and \similarity,
we consider all assertions from the codebase except for the current target assertion as candidates for in-context examples
(see \autoref{sec:eval:threats} for a discussion of this choice).

\autoref{fig:rq2_total} shows the success rate of these variants on the entire \dataset dataset,
and also splits the results by codebase.
In~\autoref{fig:rq2_total},
\similarity is the most successful variant with up to 56.6\% of lemmas verified overall,
compared to 51.7\% for \random, 46.8\% for \similaritynoplaceholder and 34.4\% for \placeholder.
Moreover, \similarity and \random dominate already for small values of $k$,
with around 40\% of lemmas solved on the first attempt.
These results confirm our two hypotheses:
\begin{enumerate*}
  \item that providing example assertions from the same codebase is beneficial (since \random outperforms \placeholder), and
  \item that providing example assertions from similar proof contexts is even more beneficial (since \similarity outperforms \random).
\end{enumerate*}
Moreover, we see that the benefits from the placeholder and the in-context examples are complementary,
since \similaritynoplaceholder does not perform as well as the two variants that combine the two techniques.

When examining the codebases individually,
we observe that \similarity consistently achieves the highest success rate across all codebases,
although for \vmc, \random performs equally well.
\cedar, with its complex assertions,
benefits the most from the in-context examples,
with at least 30\% between \placeholder and any variant with examples.
This shows that in-context examples are crucial for domain-specific lemmas with user-defined types and functions.
\paragraph{Comparison with other similarity metrics}

\begin{figure}
  \centering
  \includegraphics[width=1\linewidth]{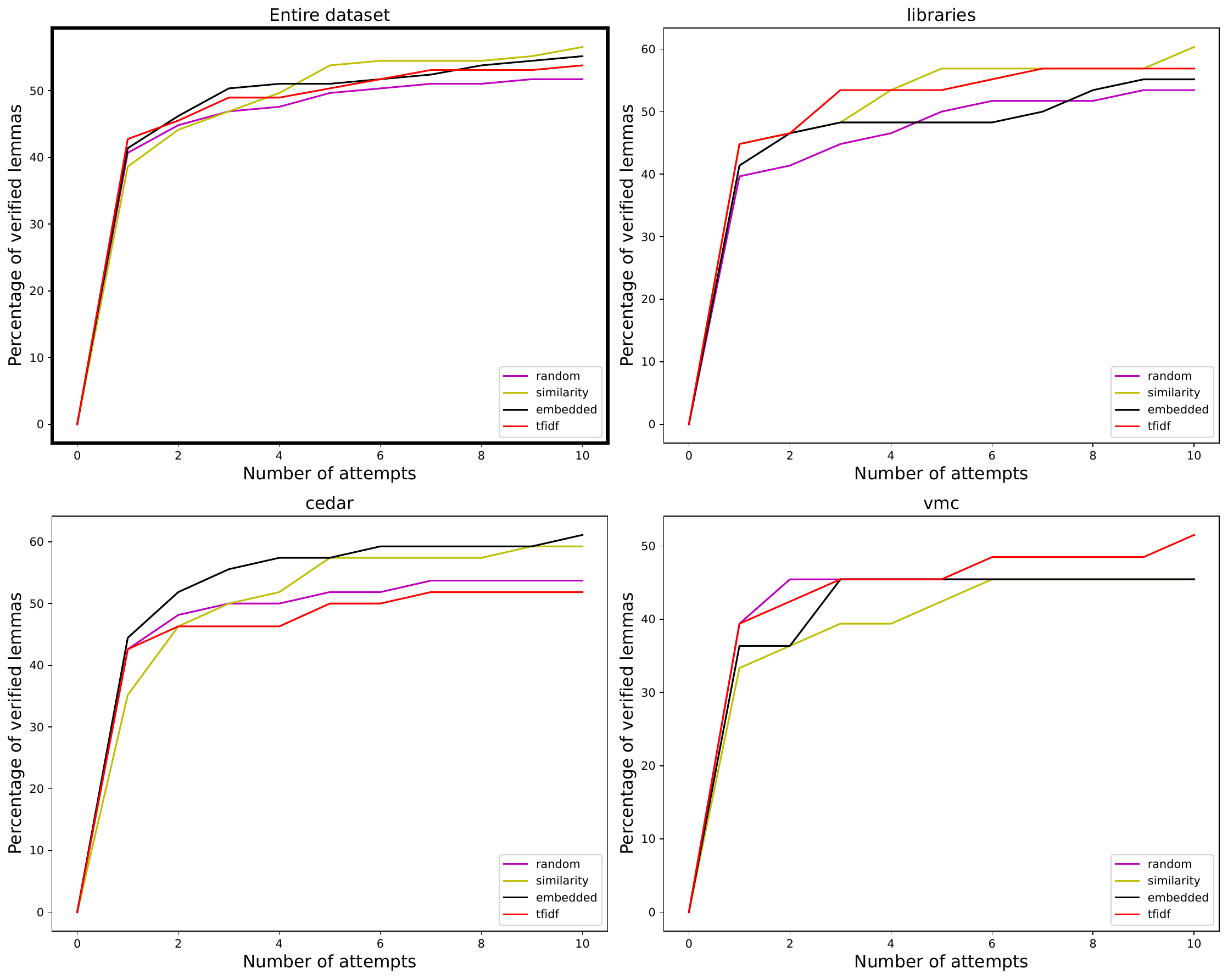}
  \caption{Percentage of verified lemmas vs inference cost for the entire \dataset dataset, and by codebase, using similarity.}
  \label{fig:rq3_bis}
\end{figure}

To evaluate the effectiveness of our \emph{proof similarity} metric,
we also compared it with two state-of-the-art similarity metrics, with the following variants:
\begin{itemize}
  \item \random: same as previous experiment, the LLM is given a placeholder and $m$ random examples from the same codebase.
  It serves as a baseline for the other similarity metrics.
  \item \similarity: our \emph{proof similarity} with a placeholder and $m$ examples from the same codebase.
  \item \embedding: the LLM is given the code with a placeholder and $m$ examples from the same codebase, selected based on neural embedding similarity.
  \item \tfidf: the LLM is given the code with a placeholder and $m$ examples from the same codebase, selected based on TF-IDF similarity.
\end{itemize}

We kept the same setup as in the previous experiment, with $m=6$,
and also including all assertions for in-context examples.

\autoref{fig:rq3_bis} shows the success rate of each similarity metric on the entire \dataset dataset,
along with the per codebase results.
In \autoref{fig:rq3_bis}, our proof \similarity is the most successful similarity metric
with 56.6\% of lemmas verified overall, compared to 55\% for \embedding, 53.6\% for \tfidf, and 51.7\% for \random.
Across the 3 codebases \similarity is the most consistent performer,
scoring 1st in \libraries, and 2nd in \cedar and tied in \vmc.
Contrastingly, \tfidf is the least consistent performer,
even underperforming \random in \cedar.
All of this suggests that while the difference between the similarity metrics is small,
the \emph{proof similarity} metric
performs at least as well as the state-of-the-art similarity metrics,
while remaining simple.
It can serve as a substitute to \embedding and \tfidf for this kind of task,
as it better captures the structure of the code.

\subsection{Failure Analysis}\label{sec:eval:failure}

To get an insight into the limitations of our techniques,
we manually analyzed a sample of tasks that could not be solved by the best configuration of \tool.
We found two sources of failure to be most prevalent,
which we describe below.

\paragraph{Function Missing from Context}

Sometimes an assertion needs to call a function defined elsewhere in the codebase.
Consider the lemma \T{SoundArith},
which states the soundness of binary arithmetic operations in the \cedar language:
\begin{dafny}
lemma SoundArith(op: BinaryOp, e1: Expr, e2: Expr, t: Type, effs: Effects)
  requires ...
  ensures ...
{
  assert TC.inferArith2(op,e1,e2,effs) == types.Ok(Type.Int);
  ...
}
\end{dafny}
The assertion in Line~4 calls the function \inlinedafny{inferArith2},
which is not mentioned anywhere else in this lemma.
Hence, if we remove this assertion and ask the LLM to synthesize it,
the LLM has no way of knowing that this function exists.
To remedy this issue,
our example selection technique is able to find similar assertions to guide the LLM, such as:
\[
  \mbox{\T{assert TC.inferContainsAnyAll(op,e1,e2,effs) == types.Ok(t')}}
\]
Given this example, the LLM actually picks up on the naming scheme for a family of related functions in this codebase,
and proposes the following assertion:
\[
  \mbox{\T{assert TC.inferArith(op,e1,e2,effs) == types.Ok(Type.Int)}}
\]
This assertion is just \emph{one character away} from the correct solution!
The function \T{inferArith} does not exist in the code;
instead, the code has two different functions---%
\inlinedafny{inferArith1} for unary operators and \inlinedafny{inferArith2} for binary operators.

To address this problem in the future,
we might either augment our prompts with a list of likely relevant functions from the codebase,
or attempt to repair synthesized assertions that call non-existent functions,
by looking for functions with similar names.

\paragraph{Function Overused in Context}

The opposite problem can also occur:
when a function is mentioned in the target lemma \emph{too many times},
the LLM might get fixated on it
and insist on using it in the assertion.
Consider the following lemma from Libraries:
\begin{dafny}
lemma LemmaFlattenConcat<T>(xs: seq<seq<T>>, ys: seq<seq<T>>)
  ensures Flatten(xs + ys) == Flatten(xs) + Flatten(ys)
  {
    if |xs| == 0 {
      assert xs + ys == ys;
    } else {
      ... // Flatten called another 6 times
    }
  }
\end{dafny}
Here \T{xs} and \T{ys} are nested sequences,
and \T{Flatten} is a helper function that flattens a nested sequence.
Importantly, this function is used a whopping \emph{nine times} in the lemma!
This is a problem when we ask the LLM to generate the assertion on Line~4,
as the LLM thinks that \T{Flatten} is important in this lemma,
and refuses to generate any assertions without it.

\subsection{Threats to Validity}\label{sec:eval:threats}

\paragraph{Internal}
The main threat to internal validity is the randomness in our experiments,
which stems from both the LLM inference and the selection of in-context examples with the \random strategy;
hence the success rates reported in \autoref{fig:rq1_total} and \autoref{fig:rq2_total} can be noisy.
Normally, such experiments need to be run multiple times,
but due to the high costs of LLM inference,
this has not been common practice in related
literature~\cite{DafnyFSE24,BaldurFSE23First}.
The reason why these results are not as noisy
as one might expect is that the success rate
is aggregated over the tasks in the dataset,
and also over the number of attempts:
the plots show whether \tool succeeded at
least once in $k$ attempts, so for higher
$k$, this measurement is more stable.

Another potential concern is \emph{data leakage},
\ie whether GPT-4o has been trained on the code
that we are evaluating it on.
Although data leakage is possible, based on the GPT-4o's knowledge cut-off date of October 2023,
we believe that it is unlikely to have a significant impact on our results,
based on the poor performance of the \baseline variant.

\paragraph{External}
The main threat to external validity is that
\tool only tries to insert one assertion
at a time, whereas in reality multiple
assertions might be needed to fix a lemma.
The single-assertion setting is representative of the scenario of fixing
brittle proofs, but might be less representative
of the scenario of writing new lemmas from scratch.
We believe that \tool can be extended to the
multi-assertion setting as follows:
\begin{enumerate}
  \item If multiple assertions are needed to fix \emph{different} verification errors,
    \tool can run independently on each error.
  \item If multiple assertions are required to fix \emph{the same} verification error,
    we can try verifying different subsets of assertions obtained from the $k$ attempts.
\end{enumerate}
The evaluation of these extensions is left for future work.
One potential issue is that the LLM sometimes uses existing
assertions from the same lemma as hints for the missing assertion,
and this might make generating multiple assertions from
scratch harder than adding a single assertion.

A related but slightly different threat
is that \tool relies on existing assertions
in the codebase to provide in-context examples.
This is only possible in a mature codebase
with many assertions, hence our example selection
technique might not work as well in the early stages
of a project (while the placeholder technique is independent
of the codebase maturity).
To study this further, we experimented with a sample of the benchmark (n=13),
where we increased the number of existing assertions in the codebase from 25\% to 100\%.
\tool's performance improves \emph{gradually}
as the number of the existing assertions increases
(see appendix \autoref{appendix:number_examples} for the results).

\tool focuses only on two prompting techniques-- \placeholder and \similarity--
as they were the most effective in our early experiments.
During development, we explored several alternative techniques, such as:
\begin{itemize}
  \item Including the complete source file from which the target lemma was extracted, which
provided excessive content that often confused the model
  \item Providing counterexamples using the Dafny option \texttt{--extract-counterexample},
but the LLM seemed to struggle to use them effectively.
  \item Attempting multiple retries for incorrect assertions within the same prompt,
which mainly helped in case of syntax errors.
\end{itemize}
We extended our investigation to more sophisticated prompting strategies,
including Chain of Thought~\cite{cot},
that breaks down
the task into a sequence of smaller and more manageable tasks (appendix \autoref{appendix:cot_prompt}).
Additionally, we tested an enhanced prompt incorporating
additional explanations on the nature and placement of assertions (appendix \autoref{appendix:enhanced_prompt}).
Ultimately, none of these techniques
provided significant improvements over \similarity and \placeholder.
This led us to focus \tool on these two techniques,
optimizing for both effectiveness and simplicity.

We evaluate using only one model (GPT-4o);
we chose this model because it is the most
recent and powerful LLM available at the
time of writing, and anecdotally we found
that smaller models like GPT-3.5 did not
perform as well on assertion generation,
while GPT-4 performed very similarly to GPT-4o.
However, the usefulness of \tool's prompting
technique might be diminished in the future,
as different and more powerful models are
developed.

Finally, we only evaluate \tool on three
Dafny codebases; we believe them to be
representative of the kind of codebases
that Dafny is used for, because they are
real-world projects and also quite diverse.
However, the effectiveness of \tool
might vary on other codebases.

\section{Related Work} \label{sec:related}

Several lines of research have looked into using machine learning in general
and LLMs in particular, to help with program verification.

\paragraph{Invariant Generation}
Static verifiers require explicit \emph{loop invariants}
to help summarize the effects of potentially unbounded number of iterations~\cite{Floyd67}.
There is rich literature~\cite{neural_inv_2021,pei2023,kamath2023Invariant,chakraborty2024invariantranking}
investigating the use of neural networks and LLMs to generate candidate loop invariants
and proposing symbolic algorithms to validate and rank them.
The reason we chose to focus on assertions instead of loop invariants in our work is practical:
we are motivated by a set of industrial verification developments,
which contain many assertions but relatively few loops.
Moreover, the symbolic algorithms used in loop invariant generation~\cite{ultimate}
are unlikely to be applicable to our dataset,
since our assertions involve reasoning about user-defined functions.

\paragraph{LLMs for Interactive Proofs}
LLMs have also been applied to generate or repair proofs
in verification languages that do not rely on SMT solvers,
such as Coq~\cite{yang2019learningprovetheoremsinteractingCoqGym,BaldurFSE23First} and Lean~\cite{leandojo,wang2023legoprover}.
The style of these ``extrinsic'' interactive
proofs, which typically comprise repetitive
application of a fixed number of proof tactics, is
very different from the ``intrinsic'' proofs
that are carried out with Dafny, which instead
requires the user to spell out non-trivial,
highly program--specific hints via assertions.
We refer the reader to a recent
survey~\cite{li2024surveydeeplearningtheorem}
that gives an overview of the many methods used
to exploit the ``interactive'' nature of such
proving with language models.

\paragraph{LLMs for Program Repair}
LLMs have also proven successful in repairing
program code~\cite{codex,khatry2023words,ProgramRepairXia2023}
demonstrating that that LLMs are more performant for single-line
code generation.
This is similar to our assertion placeholder
that constrain the LLM generation to get better
results.
However, in this case, the LLM can benefit from
the vast corpus of existing Java/Python code,
while we show that LLM-based repair can also
be applied to relatively niche verification
languages.

\paragraph{LLMs for Program Proof Synthesis}
Several groups have looked into using LLMs to \emph{synthesize}
proofs for program verification.
At a high-level, the goal here is to start with a formal specification
or contract for a particular function is supposed to do, and to then
prompt the LLM to generate the code for that function.
\citet{DafnyFSE24} and \citet{sun2024clover}
have applied LLMs to synthesize verified Dafny code
for a subset of MBPP~\cite{mbpp} benchmarks
and for textbook problems, respectively.
More recently, \cite{brandfonbrener2024VerMCTS}
present an algorithm to synthesize verified Dafny
code from specifications of about twenty tricky examples,
but where the code is repeatedly incrementally generated
and checked to constrain the ``decoding'' of the program,
thereby reducing the number of tokens needed to generate
verified code.
\fstar is another verification language that is like Dafny,
but has various other specification mechanisms including
dependent types.
\citet{chakraborty2024neuralsynthesissmtassistedprooforiented}
describe a benchmark of nearly 600K lines of open-source \fstar code,
including nearly 32K top-level \fstar contracts, and show the LLMs,
with fine-tuning can generate verified code for between 35-41\%
of the contracts.
Finally, \cite{yang2024autoverus} the authors describe AutoVerus,
a system that does the above but for a Floyd-Hoare (i.e. Dafny style)
verifier for Rust programs. Crucially, AutoVerus constructs proof-programs
via multiple LLM (agents) that individually generate the overall proof structure,
then refine it using a set of known heuristics, and then debug and repair it guided
by the verifier's error messages. Together, AutoVerus reports a very high (near 90\%)
success rate on set of 150 function contracts.
All the above show that LLMs can greatly automate the development
of short and self-contained functions, where additionally, natural
language descriptions or formal specifications of the function's contract
are available. However, such full automation is not feasible for
the large-scale verification developments we consider.

\paragraph{LLMs for Program Proof Repair}
For this reason, we choose to focus on the part
of the verification process that is most frustrating
for the user, yet amenable to automation: the generation
of assertions.
\tool is closely related to concurrently done work
described in \cite{loughridge2024dafnybench}
which curates a large dataset of Dafny programs from GitHub,
and then evaluates the effectiveness of LLMs in generating
\emph{hints} namely assertions and loop invariants needed
to verify the functions in the dataset.
The paper shows different LLMs can successfully repair nearly
60\% of procedures, but the majority of these are small standalone
functions where 26\% require no hints (invariants or assertions) at all.
In contrast, we focus on large industrial developments,
where, as we show, in context example selection is crucial
for the success of the LLMs.
The only work we are aware of that applies LLMs to real-world Dafny code
is the preliminary report~\cite{LLMDafny} that addresses a complementary problem to ours:
generating helper \emph{lemmas}.

\paragraph{In-context Example Selection}
Selecting relevant in-context examples for a given
task is an area of active research \cite{liu-etal-2022-makes,su2023selective,lu-etal-2022-fantastically,sorensen-etal-2022-information}.
~\cite{liu-etal-2022-makes,su2023selective} propose $kNN$ methods based on the cosine similarity of the neural embeddings of the examples,
while others take information-theoretic insights for prompt optimization,
like entropy \cite{lu-etal-2022-fantastically} and mutual information \cite{sorensen-etal-2022-information} estimation,
again making use of neural embeddings or logprobs.
\tool has an advantage with respect to these since our purely symbolic method is less computationally taxing.
\citet{barke-etal-2024-solving},
use clustering based on regular expressions
to select diverse examples, but their technique
is specific to spreadsheet programming.

There have also been a series of recent work concerned with example selection specifically for code generation.
\cite{mikula2023magnushammer} develops in the related area of automated theorem proving.
They use a Transformer-based, two-stage algorithm for selecting then reranking candidate premises that will be
useful in proving a goal.
\cite{leandojo} also implements a premise retrieval mechanism based on cosine similarity between embeddings of the query state and
a given candidate premise.
Both of these methods outperform the standard benchmark methods, but requires training, or fine-tuning, both embedding models.
Additionally, in these settings, the required similarity is not code-to-code, rather premise-to-proof state, so they cannot rely
on the structural similarity.

\paragraph{Code similarity}
Code similarity is a long-standing problem that has been tackled from different angles, from detecting code plagiarism,
identify code reuse and enabling code recommendation.
The methods proposed in this area can be categorized by the structure of the program they use to determine similarity.
We refer to \cite{zakerinasrabadi2023similarity} for an updated and comprehensive review on code similarity research.

Text-based techniques \cite{nicad,Chen2015DetectingAM,tukaram2019codeclone} treat source code simply as a text document.
While this enables comparison in a mostly language-agnostic way and reasonable effectiveness,
they are easily outperformed by methods that leverage the discrete structure of programs.

Learning-based methods \cite{mikula2023magnushammer,leandojo,Chochlov2022bertclone,alon2019code2vec}, such as embedding similarity, have gained prominence with the advent of Large Language Models.
Although they have shown to work on-par with the best symbolic methods, they still present a barrier to entry,
requiring expensive preprocessing and training pipelines to work effectively.

Token-based methods \cite{Prechelt2000JPlagFP,duric2013similarity,Ullah2018PlagiarismDI,Ragkhitwetsagul2019SiameseSA}, in contrast, go a step further and treat source code as a sequence of lexer tokens,
rather than individual character.
This makes them more robust to code-relevant changes such as variable or string
literal variations, while still maintaining a decent processing time, since tokenization is usually a linear time
operation.

Tree-based techniques \cite{Yu2022ASTENSBWASP,jiang2007codeclones,Tekchandani2013codeclone} consider some representation of the syntax trees of programs by using a parser, and
usually involve tree matching or tree edit distance computation.
This allows them to handle additional code-relevant modification, like block-reordering, at the expense of time-consuming
parsing and processing for large codebases and feature-rich programming languages.
Unsurprisingly, they achieve the best performance among the symbolic methods discussed.
However, their high implementation cost has limited adoption to only the most widely used programming languages,
such as C/C++ and Java.

Our Hierarchical Sequence Similarity algorithm can be seen as an intermediate between token- and tree-based similarity
algorithms, with much of the benefits of both categories.

\section{Conclusion}

We propose \tool, a tool that automatically generates assertions
to fix Dafny proofs.
On top of regular LLM techniques, \tool helps the LLM by localizing the assertion
with a \emph{placeholder} and gives qualitative examples
using our \emph{proof similarity} metric.
On our new \dataset dataset, \tool synthesizes successfully
56.6\% of all the assertions extracted from real-world code with improvements
on both the success rate and the number of attempts needed.
These numbers show that carefully crafted prompts make LLMs a suitable tool
to automate assertion generation.

\section{Data-Availability Statement}

Our artifact—containing the implementation of \tool, experiments, and results—is available
on Zenodo~\cite{emugnier_laurel_artifact} along with the \dataset dataset\cite{emugnier_dafnyGym}.
\begin{acks}
We thank our anonymous reviewers for their valuable feedback.
This work was supported by the Qualcomm Chair Endowment
and the National Science Foundation under Grants No. 1955457, 2107397, and 2422214.
\end{acks}

\bibliographystyle{ACM-Reference-Format}

\bibliography{ref}

\newpage
\onecolumn
\appendix
\section{Additional experiments}
\subsection{Varying the number of examples available for similarity}\label{appendix:number_examples}
\begin{figure}[!h]
  \centering
  \includegraphics[width=0.5\linewidth]{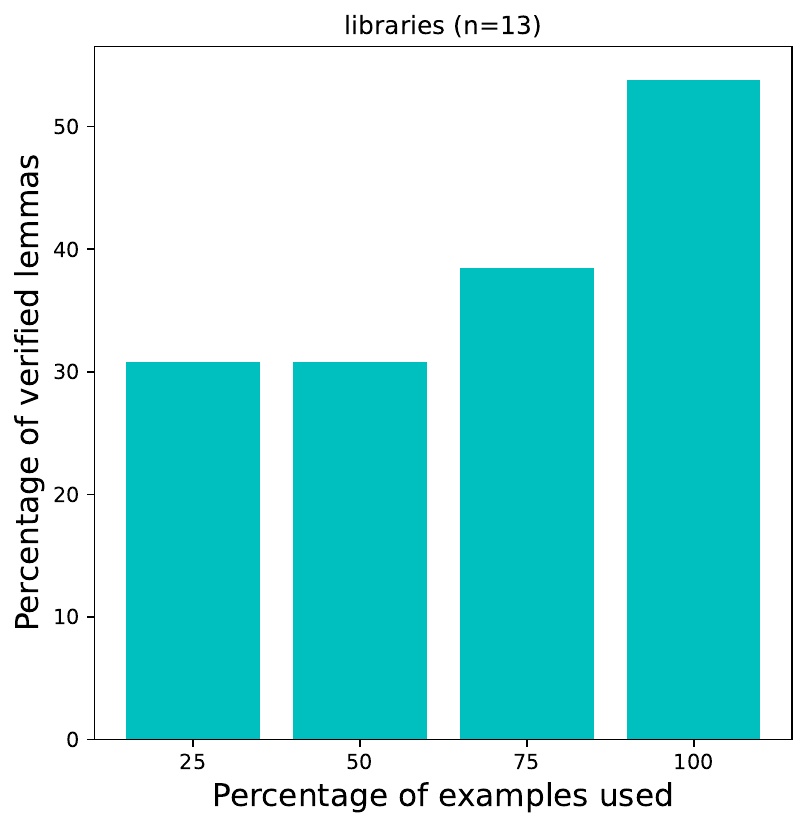}
  \caption{Percentage of verified lemmas depending on the percentage of examples available for the similarity on a sample (n=13) of the library dataset.}
  \label{fig:number_examples}
\end{figure}

\subsection{Enhanced Prompt}\label{appendix:enhanced_prompt}

We also experimented with an enhanced system prompt that includes
additional instructions on how to insert an assertion.

\begin{figure}[!h]
  \centering
  \begin{verbatim}
"You are an expert assistant on Automated Program Repair,
with proficiency in the Dafny programming language.
Your task will be to fix some incorrect Dafny code snippets.
These code snippets are incorrect because the Dafny compiler is not able to prove
the contracts that they specify, and in order to fix them you need to
insert ONE assertion statement into the provided code.
Assertions explicitly specify properties that must hold
at specific points in the program,
and introduce intermediate facts which enable the compiler
to break down proof obligations,
making them easier for the underlying SMT solver to verify.
They are indicated by the `assert` keyword and
their correct placement within the code
is crucial to ensure a correct verification of the program.
They must use only variables that are in-scope."
  \end{verbatim}
  \caption{System prompt of the enhanced prompt experiment.}
  \label{fig:enhanced_prompt_text}
\end{figure}

\begin{figure}[!h]
  \centering
  \includegraphics[width=1\linewidth]{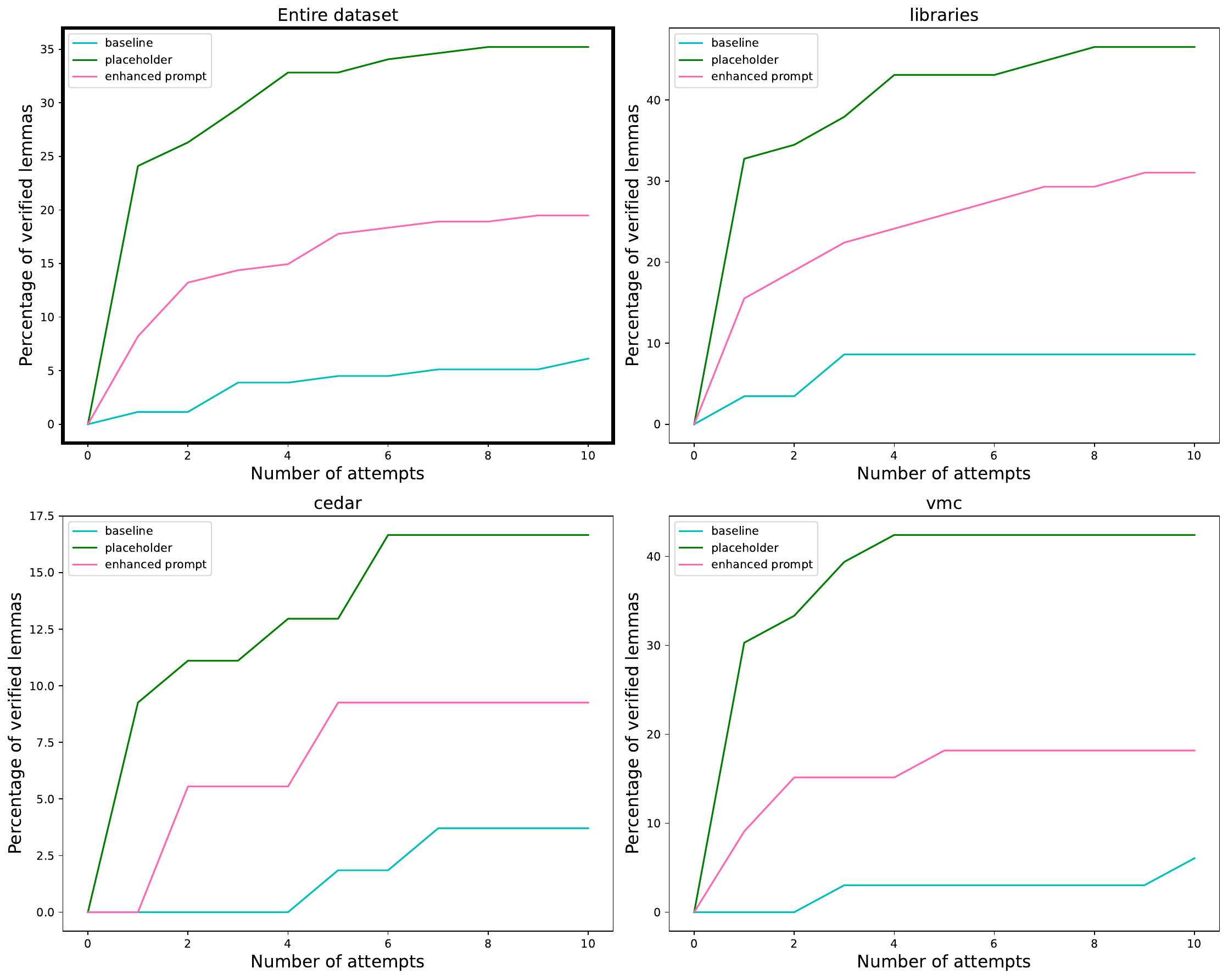}
  \caption{Percentage of verified lemmas vs inference cost for on the \dataset dataset
  with the enhanced prompt versus the original prompt.}
  \label{fig:enhanced_prompt}
\end{figure}

\subsection{Chain of Thought Prompt}\label{appendix:cot_prompt}
We also test the performance of the LLM when instructed to explain its reasoning while inserting the missing assertion,
in the spirit of chain-of-thought reasoning.
We provide precise instructions on the steps to take to fix the proof, 
which naturally align with a typical human solving strategy.
The prompt used is presented in \autoref{fig:cot_prompt}.

We tried this technique on a sample of lemmas from the \dataset dataset.
It worked well on common lemmas such as those from the \libraries codebase,
but it did not make a difference for more complicated ones.
Even when prompted in this manner, the model fails to correct the lemma,
even after 3 rounds of feedback iteration.
Instead, the model inserts irrelevant assertions,
and even hallucinates auxiliary lemmas that are not defined within the scope of the program.

\begin{figure}[!h]
  \centering
\begin{Verbatim}
You are an expert assistant on Automated Program Repair, 
with proficiency in the Dafny programming language.
Your task will be to fix some incorrect Dafny code snippets.
These code snippets are incorrect because the Dafny compiler is not able to prove 
the contracts that they specify, and in order to fix them 
you need to insert ONE assertion statement into the provided code.
Assertions explicitly specify properties that must hold at 
specific points in the program, and introduce intermediate facts 
which enable the compiler to break down proof obligations, 
making them easier for the underlying SMT solver to verify. 
They are indicated by the `assert` keyword and their correct placement 
within the code is crucial to ensure a correct verification of the program. 
They must use only variables that are in-scope.
You must produce the source code of the assertion, and after which line 
of the provided method it should be inserted.

In order to find the correct assertion, you must:
1. Understand what the lemma is proposing, and the logic that 
   it follows to prove this. This will enable you to grasp the argument 
   that is being made.
2. Find which the part where there are hypothesis missing, and establish such.
3. Return the code snippet and the line after which it must be inserted, 
   that would introduce the missing hypothesis and make the lemma verify correctly.

At the end, provide the fixed code snippet, with the new assertion inserted correctly.
\end{Verbatim}
  \caption{Prompt used in the chain-of-thought reasoning experiment.}
  \label{fig:cot_prompt}
\end{figure}
\newpage
\section{Base experiment prompt}\label{appendix:prompt}
\,
\begin{Verbatim}
# SYSTEM
"You are a Dafny formal method expert.
You will be provided with a Dafny method indicated by the delimiter <method>
that does not verify.
Your task is to insert an assertion in the <assertion> placeholder to make it verify."


# USER
Can you fix this proof by inserting one assertion in the <assertion> placeholder?
  lemma LemmaMaxOfConcat(xs: seq<int>, ys: seq<int>)
    requires 0 < |xs| && 0 < |ys|
    ensures Max(xs+ys) >= Max(xs)
    ensures Max(xs+ys) >= Max(ys)
    ensures forall i {:trigger i in [Max(xs + ys)]} :: i in xs + ys ==> Max(xs + ys) >= i
  {
    reveal Max();
    if |xs| == 1 {
    } else {
      
<assertion> Insert the assertion here </assertion>
      LemmaMaxOfConcat(xs[1..], ys);
    }
  }


# ASSISTANT
assert xs[1..] + ys == (xs + ys)[1..];


# USER
Can you fix this proof by inserting one assertion in the <assertion> placeholder?
  lemma LemmaCardinalityOfSet<T>(xs: seq<T>)
    ensures |ToSet(xs)| <= |xs|
  {
    reveal ToSet();
    if |xs| == 0 {
    } else {
      
<assertion> Insert the assertion here </assertion>
      LemmaCardinalityOfSet(DropLast(xs));
    }
  }


# ASSISTANT
assert ToSet(xs) == ToSet(DropLast(xs)) + {Last(xs)};


# USER
Can you fix this proof by inserting one assertion in the <assertion> placeholder?
  lemma LemmaFlattenLengthLeMul<T>(xs: seq<seq<T>>, j: int)
    requires forall i | 0 <= i < |xs| :: |xs[i]| <= j
    ensures |FlattenReverse(xs)| <= |xs| * j
  {
    if |xs| == 0 {
    } else {
      LemmaFlattenLengthLeMul(xs[..|xs|-1], j);
      
<assertion> Insert the assertion here </assertion>
    }
  }


# ASSISTANT
assert |FlattenReverse(xs[..|xs|-1])| <= (|xs|-1) * j;


# USER
Can you fix this proof by inserting one assertion in the <assertion> placeholder?
  lemma LemmaCardinalityOfSetNoDuplicates<T>(xs: seq<T>)
    requires HasNoDuplicates(xs)
    ensures |ToSet(xs)| == |xs|
  {
    reveal HasNoDuplicates();
    reveal ToSet();
    if |xs| == 0 {
    } else {
      LemmaCardinalityOfSetNoDuplicates(DropLast(xs));
      
<assertion> Insert the assertion here </assertion>
    }
  }


# ASSISTANT
assert ToSet(xs) == ToSet(DropLast(xs)) + {Last(xs)};


# USER
Can you fix this proof by inserting one assertion in the <assertion> placeholder?
  lemma LemmaCardinalityOfEmptySetIs0<T>(xs: seq<T>)
    ensures |ToSet(xs)| == 0 <==> |xs| == 0
  {
    reveal ToSet();
    if |xs| != 0 {
      
<assertion> Insert the assertion here </assertion>
    }
  }


# ASSISTANT
assert xs[0] in ToSet(xs);


# USER
Can you fix this proof by inserting one assertion in the <assertion> placeholder?
  lemma LemmaInvFoldLeft<A,B>(inv: (B, seq<A>) -> bool,
                              stp: (B, A, B) -> bool,
                              f: (B, A) -> B,
                              b: B,
                              xs: seq<A>)
    requires InvFoldLeft(inv, stp)
    requires forall b, a :: stp(b, a, f(b, a))
    requires inv(b, xs)
    ensures inv(FoldLeft(f, b, xs), [])
  {
    reveal FoldLeft();
    if xs == [] {
    } else {
      
<assertion> Insert the assertion here </assertion>
      LemmaInvFoldLeft(inv, stp, f, f(b, xs[0]), xs[1..]);
    }
  }


# ASSISTANT
assert [xs[0]] + xs[1..] == xs;


# USER
Can you fix this proof by inserting one assertion in the <assertion> placeholder?
 <method>   lemma LemmaMinOfConcat(xs: seq<int>, ys: seq<int>)
    requires 0 < |xs| && 0 < |ys|
    ensures Min(xs+ys) <= Min(xs)
    ensures Min(xs+ys) <= Min(ys)
    ensures forall i :: i in xs + ys ==> Min(xs + ys) <= i
  {
    reveal Min();
    if |xs| == 1 {
    } else {

<assertion> Insert assertion here </assertion>

            LemmaMinOfConcat(xs[1..], ys);
    }
  }
 </method>
\end{Verbatim}
\break

\end{document}